\documentclass[5p]{elsarticle}

\makeatletter
\def\ps@pprintTitle{%
 \let\@oddhead\@empty
 \let\@evenhead\@empty
 \def\@oddfoot{\footnotesize\itshape
       Preprint submitted to \ifx\@journal\@empty Elsevier
       \else\@journal\fi\hfill March 05, 2019}%
 \let\@evenfoot\@oddfoot}
\makeatother

\usepackage{amsmath}
\usepackage{amsfonts}
\usepackage{amssymb}
\usepackage{bm}
\usepackage{undertilde}
\usepackage{xspace}
\usepackage{siunitx}
\usepackage[colorlinks]{hyperref}

\journal{Ocean Modelling}

\biboptions{authoryear}

\defcitealias{stanley2019topobaric}{S19}

\newcommand{\di}{\partial}

\renewcommand{\d}{\mathrm{d}}

\newcommand{\dz}{\, \mathrm{d}z}
\newcommand{\norm}[1]{\lVert #1 \rVert}

\renewcommand{\deg}{\ensuremath{^{\circ}}\xspace}
\newcommand{\degn}{\ensuremath{^{\circ}\mathrm{N}}\xspace}

\newcommand{\dege}{\ensuremath{^{\circ}\mathrm{E}}\xspace}

\newcommand{\ie}{i.e.\xspace}
\newcommand{\eg}{e.g.\xspace}

\newcommand{\vs}{vs.\xspace}

\newcommand{\bu}{\bm{u}}
\newcommand{\ihat}{\bm{\hat{i}}}
\newcommand{\jhat}{\bm{\hat{j}}}
\newcommand{\khat}{\bm{\hat{k}}}

\newcommand{\rB}{\rho_B} 
\newcommand{\rBr}{\rB^{-1}} 

\DeclareRobustCommand{\surf}[1]{\utilde{#1}} 

\newcommand{\eossv}{\mathcal{V}} 
\newcommand{\eosB}{B} 
\newcommand{\sv}{\nu} 
\newcommand{\svn}{\surf{\sv}}
\newcommand{\svpn}{\surf{\sv_p}}
\newcommand{\pn}{\surf{p}}
\newcommand{\zn}{\surf{z}}
\newcommand{\sn}{\surf{S}}
\newcommand{\tn}{\surf{\theta}}
\newcommand{\rn}{\surf{\rho}}
\newcommand{\rzn}{\surf{\rho_z}}
\newcommand{\dn}{\surf{\delta}}

\newcommand{\Ps}{\surf{\Psi}} 
\newcommand{\Psp}{\Ps^{(p)}} 
\newcommand{\Psd}{\Ps^{(\delta)}} 
\newcommand{\Pszh}{\Ps^{(ZH)}} 
\newcommand{\Pscu}{\Ps^{(Cu)}} 
\newcommand{\Psmk}{\Ps^{(MK)}} 
\newcommand{\Psn}{\Ps^{(n)}} 
\newcommand{\Psom}{\Ps^{(oM)}} 
\newcommand{\Pstb}{\Ps^{(tb)}} 

\newcommand{\buzh}{\surf{\bu_g}^{(ZH)}} 
\newcommand{\bucu}{\surf{\bu_g}^{(Cu)}} 
\newcommand{\bumk}{\surf{\bu_g}^{(MK)}} 
\newcommand{\buom}{\surf{\bu_g}^{(oM)}} 
\newcommand{\butb}{\surf{\bu_g}^{(tb)}} 
\newcommand{\buzs}{\surf{\bu_g}^{(\nabla_z p)}} 

\newcommand{\butruth}{\surf{\bu_g}} 
\newcommand{\utruth}{\surf{u_g}} 

\newcommand{\svfn}{\hat{\sv}} 
\newcommand{\svpfn}{\hat{\pi}} 

\newcommand{\rfn}{\hat{\rho}} 

\newcommand{\slp}{p^{(\eta)}}

\newcommand{\isd}{\emph{in-situ} density\xspace}
\newcommand{\gsf}{GSF\xspace}
\newcommand{\gsfs}{GSFs\xspace}
\newcommand{\eqn}[1]{Eq.~#1}

\newcommand{\rv}{0} 
\newcommand{\rvm}{0} 

\newcommand{\reg}{\mathcal{R}} 

\newcommand{\xref}{180\dege}
\newcommand{\yref}{0\degn}
\newcommand{\zrefone}{-997.99} 
\newcommand{\zreftwo}{-1988.60}

\begin{document}

\begin{frontmatter}

\title{The exact geostrophic streamfunction for neutral surfaces}

\author{Geoffrey J. Stanley\fnref{fn1}}
\address{Department of Physics, University of Oxford, Oxford, OX1 3PU, United Kingdom}
\ead{g.stanley@unsw.edu.au}

\fntext[fn1]{Current address: School of Mathematics and Statistics, University of New South Wales, Sydney, NSW 2052, Australia.\\
\copyright{} 2019. This manuscript version is made available under the CC-BY-NC-ND 4.0 license http://creativecommons.org/licenses/by-nc-nd/4.0/}


\begin{abstract}
McDougall (1989) proved that neutral surfaces possess an exact geostrophic streamfunction, but its form has remained unknown. The exact geostrophic streamfunction for neutral surfaces is derived here. It involves a path integral of the specific volume along a neutral trajectory. On a neutral surface, the specific volume is a multivalued function of the pressure on the surface, $\utilde{p}$. By decomposing the neutral surface into regions where the specific volume is a single-valued function of $\utilde{p}$, the path integral is simply a sum of integrals of these single-valued functions. The regions are determined by the Reeb graph of $\utilde{p}$, and the neutral trajectory is encoded by a walk on this graph. Islands, and other holes in the neutral surface, can create cycles in the Reeb graph, causing the exact geostrophic streamfunction on a neutral surface to be multivalued.  Moreover, neutral surfaces are ill-defined in the real ocean. Hence, the topobaric geostrophic streamfunction is presented: a single-valued approximation of the exact geostrophic streamfunction for neutral surfaces, for use on any well-defined, approximately neutral surface.  Numerical tests on several approximately neutral surfaces reveal that the topobaric geostrophic streamfunction estimates the geostrophic velocity with an error that is about an order of magnitude smaller than that for any previously known geostrophic streamfunction. Also, the Montgomery potential is generalized, leading to an alternate form of the exact geostrophic streamfunction for neutral surfaces. This form is used to construct the orthobaric Montgomery potential, an easily computable geostrophic streamfunction that estimates the geostrophic velocity more accurately than any previously known geostrophic streamfunction, but less so than the topobaric geostrophic streamfunction. 
\end{abstract}


\end{frontmatter}


\section{Introduction}
\label{sec:intro}

Outside the mixed layer and bottom boundary layer and on scales larger than about $\SI{10}{m}$, the structure of oceanic flow is predominantly two-dimensional, largely confined to isosurfaces of a quasi-conservative density.
This enabled \citet{reid.lynn1971}, for example, to draw conclusions about how the deep tropical oceans are connected to the shallow polar oceans.
This method of analysis is even more powerful if, rather than just knowing the flow is within such a surface, we can describe the flow within that surface, for example by means of a geostrophic streamfunction (\gsf). 

A \gsf is the potential, in a specified surface, for the acceleration by the horizontal pressure gradient. (Hence, this quantity is sometimes called an ``acceleration potential''; nomenclature is discussed at the end of this section.)
A \gsf is not, technically, a streamfunction of the geostrophic velocity: none exists, because meridional variation of the Coriolis parameter implies the geostrophic velocity is divergent. 
Still, the geostrophic velocity (a 2D vector field) is exactly determined by the \gsf (a scalar field) and the known Coriolis parameter. 
Moreover, the geostrophic velocity is everywhere (except the equator) tangent to contours of the \gsf, which are therefore streamlines of the geostrophic velocity. 

\gsfs possess many theoretical graces, not least being their deep connection with potential vorticity. 
If a \gsf exists on isosurfaces of a materially conserved 3D variable, then
this variable gives rise to a materially conserved Ertel potential vorticity: when the momentum equations are recast with this variable as the vertical coordinate, the gradient of the \gsf exactly represents the horizontal pressure gradient acceleration, and no spurious acceleration terms are produced \citep{deszoeke2000}. 
Such 3D variables and their \gsfs are thus of prime theoretical and practical importance in layered models. 
Also, because of the divergence-free flow (namely, the geostrophic velocity multiplied by the Coriolis parameter) they provide, 
\gsfs serve as building blocks for many theoretical models \citep{stommel1948, stommel.arons1959, welander1971, rhines.young1982, luyten.pedlosky.ea1983, marshall.radko2003}
and inverse models \citep{killworth1986, cunningham2000, zika.mcdougall.ea2010},
and are useful to analyse quantities in recirculating flow, such as streamwise budgets and averages \citep{gille1997a, shuckburgh.jones.ea2009, chapman.sallee2017}.

Mapping \gsfs was first done on specific volume anomaly surfaces \citep{montgomery1938, reid1965}, on which the \citet{montgomery1937} potential is the \gsf{}. 
Later, potential density became the preferred quasi-conserved density variable, upon whose surfaces many authors analysed the geostrophic circulation by means of the Montgomery potential \citep[\eg][]{bower.rossby.ea1985, lozier.owens.ea1995, aksenov.ivanov.ea2011}. 
However, the Montgomery potential is not a \gsf for a potential density surface, and indeed none exists\citep{mcdougall1989}. 
Its use on a potential density surface can predict a geostrophic velocity that differs substantially from the true geostrophic velocity, as shown by \citet{zhang.hogg1992}, who then upgraded the Montgomery potential to minimize these errors.  We say the Montgomery potential is the exact \gsf on specific volume anomaly surfaces, but it is an inexact, or approximate, \gsf on potential density surfaces. 

Since \citet{mcdougall1987ns} defined neutral surfaces and also highlighted the problems of using potential density far from its reference pressure, another shift has occurred, towards using surfaces that are more closely aligned with the neutral tangent plane --- the plane in which fluid parcels can move adiabatically and infinitesimally without experiencing a buoyant restoring force.
Neutral trajectories and neutral surfaces are, respectively, paths and surfaces that are everywhere parallel to the local neutral tangent plane \citep{mcdougall1987ns}.
Unfortunately, non-linearity in the equation of state for seawater makes neutral trajectories path-dependent, and so neutral surfaces are formally ill-defined \citep{mcdougall.jackett1988}. However, in practice this path-dependence is small enough \citep{mcdougall.jackett1988, mcdougall.jackett2007} that we can usefully craft approximately neutral surfaces, which are well-defined and are \emph{approximately} parallel with the neutral tangent plane; examples include neutral density surfaces \citep{jackett.mcdougall1997}, orthobaric density surfaces \citep{deszoeke.springer.ea2000}, $\omega$-surfaces \citep{klocker.mcdougall.ea2009}, and topobaric surfaces \citep[][hereafter \citetalias{stanley2019topobaric}]{stanley2019topobaric}. Of course, specific volume anomaly surfaces and potential density surfaces are also approximately neutral surfaces, but ``approximately'' carries heavier emphasis.

An opportunity now presents itself: use the exact \gsf for neutral surfaces, rather than the Montgomery potential or its variants, on these approximately neutral surfaces. 
\citet{mcdougall1989} proved that there is an exact \gsf for neutral surfaces, at least in a theoretical ocean where neutral surfaces are well-defined. 
However, the analytic form of this \gsf was not given, and has remained elusive ever since. 

The key theoretical result of this paper is to derive the exact \gsf for neutral surfaces. It is found to also suffer from path-dependency, this time caused by islands and other holes in the neutral surface. 
Its ill-defined nature is overcome by the key practical result of this paper: the topobaric \gsf, which approximates the exact \gsf for neutral surfaces but is well-defined, and can be used on any approximately neutral surface. 
These key results are essentially corollaries to a theory of neutral surfaces developed in a companion paper \citepalias{stanley2019topobaric}. The root of this theory is a multivalued functional relationship between the specific volume and the pressure on neutral surfaces, and its characterisation by the \citet{reeb1946} graph. 
The necessary ideas of that theory are briefly described here, but the reader is encouraged to read the companion paper first.

Complimenting these results, the way \citet{zhang.hogg1992} upgraded the \citet{montgomery1937} potential is generalized. This generalization turns out to be none other than the exact \gsf for neutral surfaces, but presenting a different functional form. This alternate form is useful in creating an easily calculable yet highly accurate \gsf called the orthobaric Montgomery potential --- the second key practical result of this paper. 

Section~\ref{sec:theory} reviews \gsfs and neutral surfaces, then derives the exact \gsf for neutral surfaces. 
The topobaric \gsf is discussed in Section~\ref{sec:topobaric_gsf}.
The Montgomery potential and its variants are reviewed in Section~\ref{sec:montgomery}, then generalized in Section~\ref{sec:orthobaric_montgomery} to create the orthobaric Montgomery potential.
Section~\ref{sec:Boussinesq_models} discusses the model data to be used and the Boussinesq approximation.
Numerical comparisons of various \gsfs are presented in Section~\ref{sec:results},
 before summarizing in Section~\ref{sec:conclusions}.
\ref{sec:geostrf_boussinesq} provides formulas for various \gsfs in a Boussinesq ocean, and
\ref{sec:numerics} details their numerical computation. 

A final note on nomenclature is warranted,
a subject of some confusion at least since a joint letter by \citet{wexler.montgomery1941} suggested \emph{both} the names ``stream function'' and ``acceleration potential'', respectively, for the quantity here called a \gsf.
There are two issues with the name ``acceleration potential''. First, it has a specific, older definition in potential flow theory. Second, it neglects other acceleration terms; technically it should be called the ``horizontal pressure gradient acceleration potential'' in a particular surface. 
Later, the name ``Montgomery potential'' and arose, but this is best reserved for the specific function defined by \citet{montgomery1937} for use in a specific volume anomaly surface. 
``Montgomery function'' has also been used, but this name bears no direct relation to the horizontal pressure gradient acceleration, the geostrophic velocity, or surfaces other than specific volume anomaly surfaces. 
This manuscript adopts the now widely-used modification of Wexler's suggestion: ``geostrophic streamfunction'', to be thought of as a compound term with a specific definition --- not as shorthand for ``streamfunction of the geostrophic velocity'', as no such streamfunction exists.

\section{The exact geostrophic streamfunction on a neutral surface}
\label{sec:theory}

\subsection{Preliminary definitions}
The salinity, potential temperature, pressure, \isd, and specific volume are 3D scalar fields denoted by $S$, $\theta$, $p$, $\rho$, and $\sv$ respectively.\footnote{The theory presented here could equivalently use Absolute Salinity and Conservative Temperature.  Here, practical salinity and potential temperature are used to match what is used by the equation of state in the ocean model whose data shall be analysed.} They are related by $\sv = \rho^{-1} = \eossv(S,\theta,p)$ where $\eossv$ is the (inverse of the) equation of state. Also let $\sv_p = \di_p \eossv(S,\theta,p)$ be another 3D scalar field, akin to the compressibility. 
Let $\ihat$, $\jhat$, and $\khat$ be the eastward, northward, and upward unit vectors, respectively.
The depth $z$ is another 3D scalar field, giving the distance in the $\khat$ direction from each point to a reference geopotential (near the sea surface), and signed so that $z < 0$ below the reference geopotential. 
The sea surface height $\eta$ is a 2D scalar field, similar to $z$ but for the sea surface. 
Let $g$ be the gravitational acceleration.

\subsection{Background of geostrophic streamfunctions}
The geostrophic velocity $\bu_g$ is defined by the geostrophic approximation of the horizontal momentum equation,
\begin{equation}
\label{eq:geostrophy}
- \khat \times f \bu_g = \sv \nabla_z p,
\end{equation}
where $f$ is the Coriolis parameter, and 
$\nabla_z p \equiv \di_x p \vert_z \, \ihat + \di_y p \vert_z \, \jhat$
 is the gradient of pressure at constant depth $z$.
A \gsf $\Psi$ 
in a generic $r$-surface (\eg an isosurface of a 3D scalar field $r$)
is defined as that which satisfies\footnote{For some surfaces, it is useful to instead require $\nabla_r \Psi = \nabla_z p$, which gives a streamfunction for $\rho f \bu_g$ rather than $f \bu_g$ \citep{mcdougall1989}.
Also, the geopotential of the sea surface is sometimes included in the definition, \ie requiring 
$\nabla_r \Psi = \sv \nabla_z p - \nabla (g \eta)$  \citep{mcdougall.klocker2010}. As $g \eta$ can be absorbed into $\Psi$, the two definitions are essentially interchangeable. 
}
\begin{equation}
\label{eq:defstrf}
\nabla_r \Psi = \sv \nabla_z p
\end{equation}
where $\nabla_r$ is the ``projected non-orthogonal gradient'' in the $r$-surface, first introduced by \citet{starr1945}. Specifically, given a tracer $C$ (a 3D scalar field),
$\nabla_r C \equiv \di_x C \vert_r \, \ihat + \di_y C \vert_r \, \jhat$,
where
the partial derivatives are taken ``in the $r$-surface'', sampling $C$ from the $r$-surface even though the radial (vertical) position is ignored when measuring distance \citep{mcdougall.groeskamp.ea2014}. 

The first step to determine $\Psi$ is usually to transform the gradient of $p$ from one at constant depth to one in the surface, according to $\nabla_r p = \nabla_z p + \di_z p \, \nabla_r z$ \citep{starr1945}. 
This is combined with hydrostatic balance, $\di_z p = -g \sv^{-1}$, transforming \eqref{eq:defstrf} into
\begin{equation}
\label{eq:defstrf2a}
\nabla_r \Psi = \sv \nabla_r p + \nabla_r (g z).
\end{equation}
treating $g$ as constant, for simplicity of presentation.\footnote{More generally, the gravitational potential $\Phi$ can be used as the vertical coordinate instead of $z$, in which case hydrostatic balance is $\d p / \d \Phi = -\sv^{-1}$, and \eqref{eq:defstrf2a} becomes $\nabla_r \Psi = \sv \nabla_r p + \nabla_r \Phi$.
}

We use an alternative notation that captures the precise details of the projected non-orthogonal gradient, but uses a standard gradient on a transformed variable \citepalias{stanley2019topobaric}. Specifically, let $\surf{C}$ be the projection onto a perfect sphere (the centre and radius of which match those of Earth) of the restriction of a 3D field $C$ to the surface in question.\footnote{For simplicity, assume the surface in question exists at a unique depth in each water column, or not at all where it has grounded or outcropped. Were it not, simply restrict attention to a local region where this is true, and work region by region.}
With $C$ a scalar field, $\nabla \surf{C}$ can be evaluated using the standard gradient in spherical coordinates. 
The only difference between $\nabla \surf{C}$ and $\nabla_r C$ is that the former deals with a single 2D surface under consideration, whereas the latter deals with the whole 3D ocean. 
Note $C$ could instead be a vector field, such as $\bm{u_g}$. 

Considering a single $r$-surface with this notation, \eqref{eq:defstrf2a} becomes
\begin{equation}
\label{eq:defstrf2}
\nabla \Ps = \svn \nabla \pn + \nabla (g \zn).
\end{equation}
The goal is to move $\svn$ inside the gradient, expressing the RHS as the gradient of some quantity, namely $\Ps$. 
The vector field $f \surf{\bu_g}$ can then be determined entirely from the scalar field $\Ps$. 

\subsection{Background of neutral surfaces}

As discussed in Section~\ref{sec:intro}, neutral surfaces are formally ill-defined. Only under special conditions are neutral surfaces well-defined. A necessary condition is that the neutral helicity is everywhere zero \citep{mcdougall.jackett1988}.
To make theoretical progress, let us assume neutral surfaces are well-defined. Later, we will translate the theoretical results into the realistic case where this is not so.

A neutral surface is one in which specific volume variations in the surface are solely due to pressure variations in the surface. 
Mathematically, this is usually expressed as
$\nabla_n \sv = \sv_p \nabla_n p$, 
 using the projected non-orthogonal gradient in a neutral surface denoted $n$ 
 \citep{mcdougall1987ns}.
With the under-tilde notation, a neutral surface is one that satisfies
\begin{equation}
\label{eq:def_ntp_svp}
\nabla \svn = \svpn \nabla \pn.
\end{equation}
An equivalent and perhaps more familiar condition is
\begin{equation}
\label{eq:def_ntp_st}
\bm{0} = \surf{\sv_S} \nabla \sn + \surf{\sv_\theta} \nabla \tn,
\end{equation}
where $\sv_S = \di_S \eossv(S,\theta,p)$ and $\sv_\theta = \di_\theta \eossv(S,\theta,p)$. This is all we need to proceed, but for a fuller review of neutral surfaces, see \citetalias{stanley2019topobaric}.

\subsection{The exact geostrophic streamfunction on a neutral surface, locally}
\label{sec:geostrf_derive}
The neutral condition \eqref{eq:def_ntp_svp} implies the gradients, and thus the contours, of $\svn$ and $\pn$ are always aligned. Therefore, there is a functional relationship between $\svn$ and $\pn$\,:
\begin{equation}
\label{eq:svfn}
\svn = \svfn(\pn)
\end{equation}
for some function $\svfn$. In fact, $\svfn$ is a multivalued function: $\svn$ is constant on a closed contour of $\pn$, but there can be different values of $\svn$ on different contours of $\pn$ at the same pressure value. 

For now, consider a region where $\svfn$ is single-valued. 
In this region, the exact \gsf for a neutral surface is
\begin{equation}
\label{eq:geostrfexact}
\boxed{
\Psn = g \zn + \int_{p_\rv}^{\pn} \svfn (p') \, \d p'
}
\end{equation}
for some constant pressure $p_\rv$ in the domain of $\svfn$. Note $p_\rv$ provides an arbitrary additive constant to $\Psn$.
That $\Psn$ exactly satisfies \eqref{eq:defstrf2} is easily confirmed by taking its gradient using the Leibniz integral rule, then using \eqref{eq:svfn}.

\subsection{Discussion}
 
In \eqref{eq:geostrfexact}, $\zn$ is obtained by inverting hydrostatic balance to get $\di_p z = - g^{-1}  \sv$, then integrating in the local water column to get
\begin{equation}
\label{eq:zn_as_integral}
\zn = \eta - \frac{1}{g} \int_{\slp}^{\pn} \sv \, \d p,
\end{equation}
where $\slp$ is the pressure at $z = \eta$, the sea surface.
(If $\slp$ is constant, it is common, but not necessary, to set $p_\rv = \slp$.)
This transformation is standard when studying \gsfs. 
For example, multiplying \eqref{eq:zn_as_integral} by $g$ and taking the gradient in an isobaric surface (on which $\pn$ is constant) yields $\nabla (g \zn)$ on the LHS, which is the entire RHS of \eqref{eq:defstrf2} in this case. Thus 
\begin{equation}
\label{eq:dynamic_height}
\Psp = g \eta - \int_{\slp}^{\pn} \sv \, \d p
\end{equation}
is the exact \gsf in an isobaric surface; this is the well-known ``dynamic height''.\footnote{Numerical calculations of the dynamic height can be made more accurate by using the specific volume anomaly 
in the integrand of \eqref{eq:dynamic_height}; this is then called the ``dynamic height anomaly''. With double precision computing, this is no longer necessary \citep{mcdougall.klocker2010}.}
Though $\Psp$ is really just $g \zn$, the form \eqref{eq:dynamic_height} is essential because it is computable from hydrographic casts which measure the salinity and potential temperature as functions of pressure: denoting these $\check{S}(p)$ and $\check{\theta}(p)$ respectively at a given water column, the integrand in \eqref{eq:zn_as_integral} and \eqref{eq:dynamic_height} is a function of $p$ alone, namely $\eossv(\check{S}(p), \check{\theta}(p), p)$.
With model data, the actual depth of an isobaric or other such surface may be known, but this is not so for real ocean measurements. 

One might argue that \eqref{eq:geostrfexact} is not a closed form expression for $\Psn$, since $\svfn$ is unspecified. Or is it?
We are accustomed to think there is ambiguity about neutral surfaces, given their ill-defined nature in the real ocean with non-zero neutral helicity, and this ambiguity transfers to $\svfn$. 
But if we actually --- hypothetically --- have a well-defined neutral surface, then we do have the function $\svfn$. It is determined by the data $(\pn, \svn)$: a scatter plot of these would show a functional relationship around which there is precisely zero scatter.
Also, consider that $\Psp$  
has an analogous definition: it too is defined using a function of specific volume, and this function is determined from data. In this case, the data is taken vertically above or below the location in question, which is easy enough to fathom. 
 For $\Psn$, the problem lies in determining the local neutral surface, from which to get the data $(\pn, \svn)$. Assuming the neutral surface is known, $\Psn$ is the exact \gsf on that surface.

In fact, $\Psn$ from \eqref{eq:geostrfexact} is the exact \gsf on any surface for which $\svn$ is a function of $\pn$. One example is an isosurface of $\sv$, but this is very nearly a geopotential, far from neutral, and so not terribly useful. 
Dividing \eqref{eq:def_ntp_svp} by $\surf{\sv_p}$, the entire argument can be reversed, and any surface where $\pn$ is a function of $\svn$ also possesses an exact \gsf{} --- such as an isobaric surface, again far from neutral.
More generally, isosurfaces of any function $\hat{\sigma}$ that is pycnotropic (a function only of $p$ and $\sv$) possess an exact \gsf \citep{deszoeke2000}. 
This is because an isosurface of $\sigma = \hat{\sigma}(p, \sv)$ has
$0 = \nabla \surf{\sigma} = \surf{\di_\sv \sigma} \nabla \svn + \surf{\di_p \sigma} \nabla \pn$;
thus $\nabla \svn$ and $\nabla \pn$ are parallel, as in \eqref{eq:def_ntp_svp} but replacing $\surf{\sv_p}$ by
$-\surf{\di_p \sigma} \, / \, \surf{\di_\sv \sigma}$, 
and the derivation leading to \eqref{eq:geostrfexact} proceeds similarly. 
By empirically determining a pycnotropic $\hat{\sigma}$ from oceanic data, \citet{deszoeke.springer.ea2000} created orthobaric density, isosurfaces of which are approximately neutral. Orthobaric density surfaces possess an exact \gsf{}, and its form is identical to $\Psn$ in \eqref{eq:geostrfexact} --- see \eqn{2.32} of \citet{deszoeke.springer.ea2000}.  What is new about \eqref{eq:geostrfexact} is the realization that this form applies to neutral surfaces --- or rather to small regions of neutral surfaces where $\svfn$ is single-valued. 
For neutral surfaces more broadly, $\svfn$ is multivalued, and we must generalize \eqref{eq:geostrfexact} to handle this. 

\subsection{The global structure of $\Psn$}
\label{sec:geostrf_topology}

\begin{figure*}[!t]
  \centering
  \includegraphics[width=1\textwidth]
{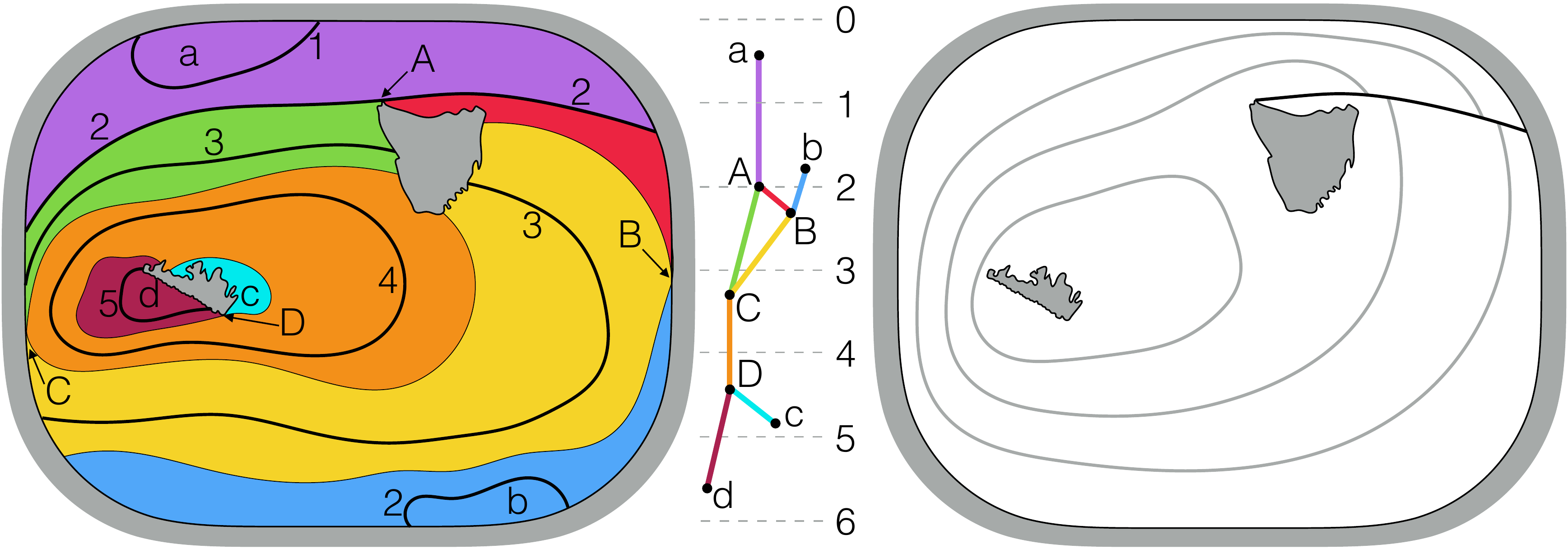}
  \caption{
  Schematic of the pressure $\pn$ on a neutral surface (left, contours in black), the exact \gsf on the same surface (right, contours in grey), and the Reeb graph of $\pn$ (centre). 
  Each node in the Reeb graph is associated with a critical point of $\pn$, and is here positioned with ordinate given by its associated critical value, and an arbitrary abscissa. 
  Each arc in the Reeb graph is associated with a region in physical space (here coloured the same as the arc) within which the multivalued function $\svfn$ is single-valued.
  Islands and other holes in the surface can create cycles in the Reeb graph, which cause the integral defining $\Psn$ to be path-dependent, hence discontinuities appear in $\Psn$ along a $\pn$ contour emanating from holes that create cycles. Here, the discontinuity is shown along a $\pn = 2$ contour (black curve in the right panel).  
  }
\label{fig:schematic}
\end{figure*}

The view so far has been local, with $\svfn$ a single-valued function. From a global perspective, $\svfn$ is multivalued: $\svn$ is constant on a contour of $\pn$, but may differ between disjoint contours of $\pn$ at the same pressure value. How this multivalued nature enters \eqref{eq:geostrfexact} must be made explicit. 

Consider an arbitrary path $\mathcal{P}$ in a neutral surface from $\bm{x}_\rv$ to $\bm{x}$. (This is a neutral trajectory.)
For simplicity of presentation, suppose the neutral surface is path-connected, so that all points $\bm{x}$ in the neutral surface are reachable from $\bm{x}_\rv$ by such a path. (If this is false, simply apply the following theory to each path-connected component separately.)
Using the gradient theorem, the global structure of $\Psn$ is
\begin{equation}
\label{eq:intPspath1}
\Psn(\bm{x}) = \Psn(\bm{x}_\rv) + \int_\mathcal{P} \nabla \Psn \cdot \d \bm{r} .
\end{equation}
Using \eqref{eq:defstrf2} and \eqref{eq:svfn} and choosing $\Psn(\bm{x}_\rv) = g \zn(\bm{x}_\rv)$ for convenience, \eqref{eq:intPspath1} becomes
\begin{equation}
\label{eq:intPspath}
\Psn = g \zn + \int_\mathcal{P} \svfn(\pn) \, \d p .
\end{equation}
It is by the neutral trajectory $\mathcal{P}$ that the multivalued nature of $\svfn$ enters $\Psn$ in \eqref{eq:intPspath}, whereas this was less clear from \eqref{eq:geostrfexact}. 
Of course, to use the gradient theorem we assumed $\Psn$ is well-defined. But is $\Psn(\bm{x})$ independent of the choice of the path $\mathcal{P}$ from $\bm{x}_\rv$ to $\bm{x}$? 
To answer this, the Reeb graph is introduced.

A contour of $\pn$ is a connected component of the level set of $\pn$ at a specified pressure value. A level set can be the disjoint union of multiple contours. The Reeb graph of $\pn$ contracts each contour of $\pn$ to a single point. 
The essence of the resulting object can be represented as a graph, a collection of $N$ nodes and $A$ arcs between pairs of nodes.
Each node $n$ corresponds to a critical point of $\pn$, denoted $\bm{x}_n$, at which point the critical value is denoted $p_n$; that is, $p_n = \pn(\bm{x}_n)$. Leaf nodes correspond to extrema of $\pn$, while internal nodes correspond to saddle points of $\pn$.
Each arc $a$ corresponds to a geographic region $\reg_a$ within which there is precisely one $\pn$ contour per pressure value.
In this way, the Reeb graph partitions space into geographic regions within each of which the multivalued function $\svfn$ is actually single-valued.
Restricting the multivalued $\svfn$ to $\reg_a$ gives a single-valued function, denoted $\svfn_a$ and called a branch of $\svfn$. 
An example is shown in the left and centre panels of Fig.~\ref{fig:schematic}. 
For oceanographic data, many of the associated regions are the shape of mesoscale eddies, which form closed $\pn$ contours. 
 
Now re-consider the neutral trajectory $\mathcal{P}$. Each point on $\mathcal{P}$ is part of a contour of $\pn$ that is contracted to some point on the Reeb graph. In this way, $\mathcal{P}$ corresponds to a walk\footnote{In graph theory, a walk is an alternating sequence of nodes and arcs where each node incident upon each arc adjacent to it in the sequence. Moreover, the sequence must start and end with nodes, but this is relaxed here.} through the Reeb graph, alternately moving along arcs and passing through nodes, in a sequence
$a_0$, $n_1$, $a_1$, ..., $n_\mathcal{J}$, $a_\mathcal{J}$.
Moreover, $\bm{x}_\rv$ is in $\reg_{a_0}$, so $\svn(\bm{x}_\rv) = \svfn_{a_0}(p_\rv)$, where $p_\rv = \pn(\bm{x}_\rv)$.
Similarly, $\bm{x}$ is in $\reg_{a_\mathcal{J}}$, so $\svn(\bm{x}) = \svfn_{a_\mathcal{J}}\big(\pn(\bm{x}) \big)$.
Thus, the path integral in \eqref{eq:intPspath} becomes a ``graph integral'', 
\begin{alignat}{3}
\label{eq:geostrfexact2}
\Psn = g \zn \quad
&+& &\int_{p_\rv}^{p_{n_1}} &&\svfn_{a_0}(p') \, \d p' \nonumber \\
&+& \quad \sum_{j=1}^{\mathcal{J}-1} &\int_{p_{n_j}}^{p_{n_{j+1}}} &&\svfn_{a_j}(p') \, \d p' \nonumber \\
&+& &\int_{p_{n_{\mathcal{J}}}}^{\pn} &&\svfn_{a_{\mathcal{J}}}(p') \, \d p'. 
\end{alignat}
This reduces to \eqref{eq:geostrfexact} when $\mathcal{J} = 0$, as happens when $\bm{x}$ and $\bm{x}_\rv$ are in the same region $\reg_a$, and $\mathcal{P}$ stays within this region. 
Note that \eqref{eq:geostrfexact2} is analogous to \eqn{20} of \citetalias{stanley2019topobaric}, in which $\svfn$ is obtained by integrating a multivalued function $\svpfn$ that satisfies $\svpn = \svpfn(\pn)$.\footnote{\citetalias{stanley2019topobaric} presents the theory of topobaric surfaces using \isd rather than specific volume, but it is trivial to use the latter instead.}

What are the circumstances that guarantee $\Psn$ is well-defined? If all paths from $\bm{x}_\rv$ to $\bm{x}$ are equivalent to the same walk through the Reeb graph, then \eqref{eq:geostrfexact2} guarantees path-independence of $\Psn$. Of course, a path could backtrack on itself, so that its walk has repeated nodes and differs from the walk of another path between the same end points. Fortunately, such backtracking is perfectly cancelled by the integrals in \eqref{eq:geostrfexact2}. Thus, $\Psn$ is well-defined if, for every pair of nodes in the Reeb graph with a walk between them, there is a unique walk between them having no repeated nodes. 

In other words, $\Psn$ is well-defined if the Reeb graph of $\pn$ is a tree, \ie it contains no cycles. (If the neutral surface is not path-connected, $\Psn$ is well-defined if the Reeb graph is a forest, \ie a collection of trees.) A cycle is a walk in the graph that starts and ends at the same node, and contains no repeated nodes or arcs, aside from the first and last node.  Cycles can arise in the Reeb graph of $\pn$ when the surface has holes, such as made by islands, seamounts, and other places where the surface grounds or outcrops. 
However, not every hole in the surface produces a cycle in the Reeb graph. For example, if a $\pn$ contour encloses a single hole, that hole does not produce a cycle \citepalias{stanley2019topobaric}. The small island in Fig.~\ref{fig:schematic} does not produce a cycle, whereas the big island does.  
Thus, one cannot assess whether $\Psn$ is well-defined or not based purely on the existence of holes in the neutral surface: the Reeb graph is needed.

There is a second, special circumstance in which $\Psn$ is well-defined, even when the Reeb graph possesses cycles. Let $n_1$, $a_1$, $n_2$, $a_2$, $...$, $a_{\mathcal{J}-1}$, $n_\mathcal{J} = n_1$ be a cycle (all $a_j$ are distinct), 
and let $\bm{x}_\rv = \bm{x} = \bm{x}_{n_1}$ without loss of generality. 
If $\Psn$ is well-defined, then \eqref{eq:geostrfexact2} along this cycle reduces to
\begin{equation}
\label{eq:island}
0 = \sum_{j=1}^{\mathcal{J}-1} \int_{p_{n_j}}^{p_{n_{j+1}}} \svfn_{a_j}(p') \, \d p'.
\end{equation}
Noting this cycle was arbitrary, \eqref{eq:island} must hold for every such cycle. If it does, then all paths between a given pair of endpoints yield the same value for $\Psn$, despite the paths potentially taking distinct walks. Thus, $\Psn$ is well-defined if and only if \eqref{eq:island} holds for every cycle in the Reeb graph of $\pn$ (which is trivially true when the graph is a tree).\footnote{In fact, this need only hold on a subset of cycles, namely a cycle basis; see \citetalias{stanley2019topobaric} for details.}

In general, the Reeb graph has cycles and there is no reason for \eqref{eq:island} to hold, so $\Psn$ is ill-defined.
Indeed, $\Psn$ is a multivalued function of geographic location, because it can be evaluated using different path integrals that loop, any number of times, around holes in the neutral surface. 
Pictured as a surface with radial (vertical) coordinate given by its value, $\Psn$ qualitatively resembles a multistorey car park, with interior ramps around holes in the neutral surface
(Fig.~2 of \citetalias{stanley2019topobaric} illustrates this, while simultaneously illustrating a similar phenomenon about neutral surfaces themselves). The ``pitch'' of these interior ramps is the (generally non-zero) RHS of \eqref{eq:island}. 
A map of $\Psn$ --- that chooses a single value of $\Psn$ for each geographic location --- will exhibit discontinuities that emanate from islands and other holes in the neutral surface, as illustrated in Fig.~\ref{fig:schematic}. 
The proof by \citet{mcdougall1989} for the existence of $\Psn$ was local in scope, so this issue could not have been foreseen.

The multivalued nature of $\Psn$ does not, of course, mean the geostrophic velocity is multivalued or discontinuous.
The geostrophic velocity is determined not by $\Psn$, but by its gradient; the path-ambiguity of $\Psn$ adds a constant offset to each ``storey'' of $\Psn$ and so does not change its gradient (evaluated using a consistent ``storey''). 
Indeed, $\nabla \Psn$ is determined by the gradient of \eqref{eq:intPspath} or \eqref{eq:geostrfexact2}, which just reverts to the unique value given by \eqref{eq:defstrf2} upon using \eqref{eq:svfn}.
So, if one desires only the geostrophic velocity, simply use \eqref{eq:defstrf2}. But often, more advanced analyses (as discussed in Section~\ref{sec:intro}) require the \gsf itself --- and require it to be well-defined, without discontinuities emanating from holes. Remedying these discontinuities is the goal for the next section.

\section{The topobaric geostrophic streamfunction}
\label{sec:topobaric_gsf}

From the previous section's theoretical results, a practical idea emerges for a well-defined \gsf that approximates $\Psn$ and is useful on any approximately neutral surface. This approximation is called the topobaric \gsf and denoted $\Pstb$ because it is built upon a topological analysis of the pressure on an approximately neutral surface. 

Given an approximately neutral surface on which the pressure is $\pn$, $\Pstb$ is constructed as follows. 
First, calculate the Reeb graph of $\pn$. 
Then, obtain a multivalued function $\svfn$ that satisfies \eqref{eq:island} for each cycle in the Reeb graph, and whose branches $\svfn_a$ approximately satisfy \eqref{eq:svfn} in each region, \ie $\svn(\bm{x}) \approx \svfn_a \big( \pn(\bm{x}) \big)$ for all $\bm{x}$ in $\reg_a$.
Then, obtain $\Pstb$ by integrating $\svfn$ according to the RHS of \eqref{eq:geostrfexact2}. 
Because $\svfn$ satisfies the cycle constraints \eqref{eq:island}, $\Pstb$ is well-defined.

When the surface in question is a topobaric surface, $\svfn$ is given \emph{a priori}. Otherwise, $\svfn$ must be empirically determined. These two cases are discussed next. 

\subsection{Use on topobaric surfaces}

It should be emphasized now that $\Pstb$ is \emph{not} the exact \gsf on a topobaric surface. 
A topobaric surface \citepalias{stanley2019topobaric} is a well-defined surface
that satisfies \eqref{eq:svfn} not just approximately, but exactly, and $\svfn$ is obtained by integrating another multivalued function $\svpfn$ which approximately satisfies $\svpn \approx \svpfn(\pn)$. (To ensure topobaric surfaces are well-defined, $\svpfn$ satisfies cycle constraints identical to \eqref{eq:island} but the integrands use $\svpfn$ rather than $\svfn$ --- see \eqn{21} of \citetalias{stanley2019topobaric}.) The resulting $\svfn$ does not, in general, satisfy the cycle constraints \eqref{eq:island}. Like true neutral surfaces, topobaric surfaces do possess an exact \gsf, but it is the multivalued $\Psn$, not the well-defined $\Pstb$.

However, $\Pstb$ is the exact \gsf on a \emph{modified} topobaric surface, which is a topobaric surface in which $\svfn$ satisfies the cycle constraints \eqref{eq:island}. 
The methods of \citetalias{stanley2019topobaric} to create topobaric surfaces are easily extended to create modified topobaric surfaces, simply by adding the cycle constraints \eqref{eq:island} when empirically fitting $\svpfn$.  
With $\svfn$ expressed in terms of $\svpfn$ (according to \eqn{20} of \citetalias{stanley2019topobaric}, analogous to \eqref{eq:geostrfexact2} here), some algebra reduces \eqref{eq:island} to
\begin{align}
\label{eq:island2}
0 = \sum_{j=1}^{\mathcal{J}-1} \Biggl( &(p_{n_{j+1}} - p_{n_j}) \sum_{k=1}^{j-1} \int_{p_{n_k}}^{p_{n_{k+1}}} \svpfn_{a_k}(p') \, \d p' \nonumber \\
&+ \int_{p_{n_j}}^{p_{n_{j+1}}}  \int_{p_{n_j}}^{p'} \svpfn_{a_j}(p'') \, \d p''  \, \d p' \Biggr),
\end{align}
which must be satisfied for each cycle $n_1$, $a_1$, $...$, $a_{\mathcal{J}-1}$, $n_\mathcal{J} = n_1$  in the Reeb graph of $\pn$. 
These additional constraints tend to make modified topobaric surfaces slightly less neutral than regular topobaric surfaces, but their exact \gsf, $\Pstb$, is well-defined. 

A goal for Section~\ref{sec:results} is to demonstrate, numerically, that $\Pstb$ is exact on modified topobaric surfaces.

\subsection{Use on other approximately neutral surfaces}

Now consider approximately neutral surfaces more generally. No exact \gsf exists for these surfaces in general, but $\Pstb$ can be constructed as an approximate \gsf. Now, $\svfn$ is not given \emph{a priori}, and \eqref{eq:svfn} is only approximate. Nonetheless, $\svfn$ can be empirically fit, then integrated to obtain $\Pstb$.
This is exactly analogous to how, for topobaric surfaces, $\svpfn$ is empirically fit, then integrated to obtain $\svfn$. Hence, the method is only summarized here; see \citetalias{stanley2019topobaric} for further details.

First, a reference location $\bm{x}_\rv$ is chosen: following \citetalias{stanley2019topobaric}, $\bm{x}_\rv = (\xref, \yref)$ by default. This is used to select a reference salinity $S_\rv = \sn(\bm{x}_\rv)$ and a reference potential temperature $\theta_\rv = \tn(\bm{x}_\rv)$.

Second, the Reeb graph of $\pn$ is calculated by the algorithm of \citet{doraiswamy.natarajan2013}. 

Third, the branches of $\svfn$ are empirically fit in each region. 
Given an arc $a$, denote the two nodes incident to $a$ as $\ell_a$ and $h_a$, with the critical values of $\pn$ at these nodes satisfying $p_{\ell_a} < p_{h_a}$.
Using a simple functional form
\begin{equation}
\label{eq:svfn_form}
\svfn_a(p') = K_a + L_a \, (p' - p_{\ell_a}) + \eossv(S_\rv, \theta_\rv, p'),
\end{equation}
the unknown constants $K_a$ and $L_a$ are determined by fitting $K_a + L_a (\pn - p_{\ell_a})$ to the specific volume anomaly on the surface, $\dn = \svn - \eossv(S_\rv, \theta_\rv, \pn)$, using ordinary least squares and restricting data to the region $\reg_a$.
Including $\eossv(S_\rv, \theta_\rv, p')$ in \eqref{eq:svfn_form} helps to capture some of the non-linearity in the equation of state. 
Each branch is fit independently, except for those whose arcs exist on cycles
 of the Reeb graph: these branches are also fit by ordinary least squares, but as a coupled problem subject to the cycle constraints \eqref{eq:island}.
For truly neutral surfaces and (modified or regular) topobaric surfaces, $\svfn$ meets continuously at the pressure saddles \citepalias{stanley2019topobaric}; however, we are using $\Pstb$ on other surfaces, so $\svfn$ is allowed to meet discontinuously at the pressure saddles. This is analogous to how topobaric surfaces allow the branches of $\svpfn$ to meet discontinuously at the pressure saddles.

Reassuringly, any such discontinuities are eliminated in the fourth step, which obtains $\Pstb$ by integrating $\svfn$ according to the RHS of \eqref{eq:geostrfexact2}. Practically, this is done by a breadth-first search in the Reeb graph:
at each step a node $m$ is discovered that is adjacent to a previously discovered node $n$ (to initialize the search, one chosen node is marked as discovered), and the arc $a$ that is incident to both $m$ and $n$ is determined; then $\Pstb$ in the region $\reg_a$ is obtained from \eqref{eq:geostrfexact} using the local branch $\svfn_a$ and using $p_0 = p_n$, so that $\Pstb$ matches continuously at $\bm{x}_n$. 
At the end of the breadth-first search, every arc has been processed except one arc $a$ per cycle. 
Obtain $\Pstb$ in each $\reg_a$ as above, using $p_0 = p_{l_a}$. 
The cycle constraints \eqref{eq:island} ensure the result is identical had we instead used $p_0 = p_{h_a}$.

\section{The Montgomery potential and variants}
\label{sec:montgomery}

We now shift gears, reviewing the \citet{montgomery1937} potential and its variants, as preparation for deriving another new \gsf in Section~\ref{sec:orthobaric_montgomery}.
The Montgomery potential is the exact \gsf on specific volume anomaly surfaces, \ie isosurfaces of the specific volume anomaly
\begin{equation}
\label{eq:sva}
\delta = \sv - \sv_\rvm,
\end{equation}
where $\sv_\rvm = \eossv(S_\rvm, \theta_\rvm, p)$ is the specific volume of a fluid parcel with reference salinity $S_\rvm$ and reference potential temperature $\theta_\rvm$, at the local pressure. 
 
To derive the Montgomery potential, add and subtract $\surf{\sv_\rvm} \nabla \pn$ from the RHS of \eqref{eq:defstrf2}. The subtracted term combines with $\svn \nabla \pn$ to give $\dn \nabla \pn$, upon which the product rule is used; the added term has the Leibniz integral rule applied to it. The result is
\begin{equation}
\label{eq:montgomery}
\nabla \Psd = \nabla \left( \dn \, \pn + \int_{P}^{\pn} \sv_\rvm \, \d p + g \zn \right) - \pn \nabla \dn,
\end{equation}
for some constant pressure $P$. 
On an isosurface of $\delta$, the last term is zero, so $\Psd$ is just the expression in parentheses. As before, the arbitrary $P$ provides an arbitrary additive constant to $\Psd$, and $\zn$ is given by \eqref{eq:zn_as_integral}. If $\slp$ in \eqref{eq:zn_as_integral} is constant, we may choose $P = \slp$, so that 
\begin{equation}
\label{eq:montgomery_form_change}
\int_{P}^{\pn} \sv_\rvm \, \d p + g \zn
= - \int_{\slp}^{\pn} \delta \, \d p + g \eta. 
\end{equation}
Substituting \eqref{eq:montgomery_form_change} into \eqref{eq:montgomery} gives a more familiar expression for $\Psd$. 
We shall continue with the more general form for non-constant $\slp$, but if $\slp$ is constant, one can also substitute \eqref{eq:montgomery_form_change} in any of \eqref{eq:ZH}, \eqref{eq:MK}, \eqref{eq:CU}, and \eqref{eq:orthobaric_montgomery} below.

On surfaces other than specific volume anomaly surfaces, $\dn$ is non-constant, so $\pn \nabla \dn$ in \eqref{eq:montgomery} is non-zero; this creates an error in estimating the geostrophic velocity by use of $\Psd$. 
To minimize this error, \citet{zhang.hogg1992} subtracted a constant pressure $p_\rvm$ from $\pn$ in \eqref{eq:defstrf2}, and percolated $p_\rvm$ through the above derivation. Equivalently, add and subtract $\nabla (\dn \, p_\rvm)$ from \eqref{eq:montgomery}, to obtain
\begin{align}
\label{eq:ZH}
\nabla \Pszh = \
&\nabla \left( \dn (\pn - p_\rvm) + \int_{P}^{\pn} \sv_\rvm \, \d p + g \zn \right) \nonumber \\
&- (\pn - p_\rvm) \nabla \dn,
\end{align}
having used $\dn \nabla p_\rvm = \bm{0}$. 
When the surface in question is not an isosurface of $\delta$, the error in using $\Pszh$ can be made less than that of $\Psd$ by choosing $p_\rvm$ to reduce the prefactor $(\pn - p_\rvm)$. \citet{zhang.hogg1992} took $p_\rvm$ as the mean of $\pn$, which minimizes the root-mean-square of $(\pn - p_\rvm)$. 
Traditionally, $(S_\rvm, \theta_\rvm)$ is taken as ($\SI{35}{psu}$, $0\deg\mathrm{C}$), following \citet{montgomery1937}.

\citet{mcdougall.klocker2010} augmented $\Pszh$ (which is designed for $\delta$-surfaces) with additional information appropriate to neutral surfaces, obtaining 
\begin{align}
\label{eq:MK}
\Psmk 
= \ &\frac{1}{2} (\pn - p_\rvm) \dn 
+ \int_{P}^{\pn} \sv_\rvm \, \d p + g \zn \nonumber \\
& -\frac{1}{12} \frac{T_b}{\rn} (\tn - \theta_\rvm) (\pn - p_\rvm)^2,
\end{align}
which is their \eqn{62} re-expressed using \eqref{eq:montgomery_form_change} and with $g \eta$ added.
To derive \eqref{eq:MK}, $T_b / \rn$ is treated as constant, namely $\SI{2.7e-15}{K^{-1}.Pa^{-2}.m^2.s^{-2}}$.

The choice of reference values $p_\rvm$, $S_\rvm$, and $\theta_\rvm$ does affect the errors.  
For $\Pszh$, one would ideally choose $p_\rvm$ in tandem with $S_\rvm$ and $\theta_\rvm$ to minimize $| (\pn - p_\rvm) \nabla \dn |$. Such optimization efforts will not be pursued here, as there are greater gains to be had by using other \gsfs. Instead, we follow the suggestion by \citet{mcdougall.klocker2010} to take 
$p_\rvm = \pn(\bm{x}_\rvm)$, 
$S_\rvm = \sn(\bm{x}_\rvm)$, and
$\theta_\rvm = \tn(\bm{x}_\rvm)$ with 
$\bm{x}_\rvm$ in the equatorial Pacific; for consistency with Section~\ref{sec:topobaric_gsf}, we take $\bm{x}_\rvm = (\xref, \yref)$.
(This will indeed give $p_\rvm$ quite close to the mean of $\pn$ over the whole surface.)

Finally, the \citet{cunningham2000} \gsf is \citep[following \eqn{25} of][]{mcdougall.klocker2010} 
\begin{equation}
\label{eq:CU}
\Pscu
= \int_P^{\pn} \eossv(\sn, \tn, p') \, \d p' + g \zn.
\end{equation}
Note that $\Pscu$ is similar to $\Psd$ but uses water-column specific ``reference'' values, namely $\sn$ and $\tn$. This re-defines $\delta$ as $\sv - \eossv(\sn, \tn, p)$, leading to $\dn = \svn - \eossv(\sn,\tn,\pn) = 0$. However, a new type of error is created due to gradients of the ``reference'' values: by the Leibniz integral rule, 
\begin{equation}
\label{eq:CU_grad}
\nabla \Pscu = \svn \nabla \pn + \nabla (g \zn) + \int_P^{\pn} \nabla_{p'} \eossv(\sn,\tn,p') \, \d p'.
\end{equation}
The last term is the error, which can be minimized by careful choice of $P$.  For consistency, we again use $\bm{x}_\rv = (\xref, \yref)$ and set $P = \pn(\bm{x}_\rv)$.

\section{The orthobaric Montgomery potential}
\label{sec:orthobaric_montgomery}

We now generalize the progress that \citet{zhang.hogg1992} made on the \citet{montgomery1937} \gsf. 
Let $\hat{p}$ be a function of $\delta$. 
Add and subtract $\hat{p}(\dn) \nabla \dn$ from \eqref{eq:montgomery}, using Leibniz's integral rule on the subtracted term, to obtain
\begin{align}
\label{eq:orthobaric_montgomery}
\nabla \Psom = \
&\nabla \left( \dn \, \pn - \int_{\Delta}^{\dn} \hat{p}(\delta') \, \d \delta' + \int_P^{\pn} \sv_\rvm \, \d p + g \zn \right) \nonumber \\
& - \big( \pn - \hat{p}(\dn) \big) \nabla \dn,
\end{align}
for some constant $\Delta$. 
The expression in large parentheses is a generalization of the Montgomery potential.

In a surprising twist, $\Psom$ has merely re-expressed $\Psn$ in the form of a Montgomery potential, as follows:
\begin{alignat}{3}
\label{eq:EM_equiv_OB}
\Psn 
&= g \zn &+& \int_P^{\pn} \svfn(p') \, \d p' && \nonumber \\
&= g \zn &+& \int_P^{\pn} \sv_\rvm \, \d p &&+ \int_P^{\pn} \hat{\delta}(p') \, \d p'  \nonumber \\
&= g \zn  &+& \int_P^{\pn} \sv_\rvm \, \d p \ &&+ \dn \, \pn - P \Delta - \int_\Delta^{\dn} \hat{p} (\delta') \, \d \delta' \nonumber \\
&= \Psom && - P \Delta. && 
\end{alignat}
In the second equality, $\hat{\delta}$ is defined by $\svfn$ and the reference profile $\sv_\rvm$: specifically, $\hat{\delta}(p') = \svfn(p') - \eossv(S_\rvm, \theta_\rvm, p')$. 
The third equality uses Laisant's inverse integral rule, with $\hat{p}$ the inverse function of $\hat{\delta}$. (If the latter is not invertible, first replace the integral of $\hat{\delta}$ with a sum of integrals over restricted domains on which $\hat{\delta}$ is invertible.) 
Finally, $P \Delta$ is an arbitrary constant, which can be ignored. 

Perhaps this equivalence between $\Psom$ and $\Psn$ should not be a surprise. 
The error in using $\Psom$ to estimate the geostrophic velocity can be made small by choosing $\hat{p}$ so that $\hat{p}(\dn)$ stays close to $\pn$.
This is eminently possible if the surface in question is a neutral surface: then $\pn$ and $\svn$ are perfectly functionally related, hence so too are $\pn$ and $\dn = \svn - \eossv(S_\rvm, \theta_\rvm, \pn)$. This gives $\hat{p}$ such that $\pn - \hat{p}(\dn) = 0$ identically, making $\Psom$ exact on neutral surfaces. 
Of course, this requires $\hat{p}$ to be a multivalued function of $\delta$, with branches defined on the Reeb graph of $\dn$. 
In this case, $\Psom$ re-expresses $\Psn$ from \eqref{eq:geostrfexact2} in the form of the Montgomery potential, and we would call it the topobaric Montgomery potential. However, this is little different from the topobaric \gsf since the Reeb graph ensures all branches are single-valued. 

Instead, the goal here is simplicity, so we take $\hat{p}$ as a single-valued function of $\delta$, and call $\Psom$ the orthobaric Montgomery potential. This re-expresses the orthobaric geostrophic streamfunction --- $\Psn$ from \eqref{eq:geostrfexact} with a single-valued $\svfn$ --- in the form of a Montgomery potential. 
The function $\hat{p}$ is empirically fit over an entire (connected part of the) surface; as such, $\Psom$ is only an approximate \gsf on surfaces other than specific volume anomaly surfaces. 

Numerical tests have persuaded the author that the orthobaric Montgomery potential 
is typically more accurate than the orthobaric \gsf{}, 
despite their mathematical equivalence. 
 This is because $\svn$ (and $\dn$) can differ substantially between the Arctic and the Southern Ocean, say, even at the same pressure: it is important to handle $\svfn$ and $\hat{\delta}$ as multivalued. On the other hand, $\hat{p}$ can more closely approximate $\pn$ as a single-valued function of $\dn$, provided the reference values $S_\rvm$ and $\theta_\rvm$ are well-chosen. 
(Think of $\pn$ as a quadratic function of $\dn$, nicely single-valued, whereas $\dn$ is like a square-root of $\pn$, multivalued. Data illustrating this point will be shown in Section~\ref{sec:results_maps}.)
To separate $\dn$ in the Arctic from $\dn$ in the Southern Ocean, we will choose $S_\rvm = \sn(\bm{x}_{\mathrm{deep}})$ and $\theta_\rvm = \tn(\bm{x}_{\mathrm{deep}})$, where $\bm{x}_{\mathrm{deep}}$ is such that $\pn(\bm{x}_{\mathrm{deep}})$ is a maximum. 
Other values for $(S_\rvm, \theta_\rvm)$ may be chosen, but this simple method seems to yield good results on a variety of surfaces. 
Then, $\hat{p}$ will be obtained by fitting $\surf{p}$ to $\dn$ as a cubic spline with 12 pieces. Again, this form seems to robustly yield good results, but other choices are possible.

\section{Gridded data and Boussinesq ocean models}
\label{sec:Boussinesq_models}

Numeric calculations in the next section will use ECCO2 data \citep{menemenlis.hill.ea2005} on 22--24 December 2002. 
The seawater Boussinesq approximation \citep{young2010}, employed by ECCO2, changes the above theory in two important, yet easy-to-handle ways.
First, Boussinesq ocean models swap the \emph{in-situ} density $\rho$ for a constant reference density $\rB$ in the horizontal momentum equations, so 
\eqref{eq:geostrophy}--\eqref{eq:defstrf2} become 
\begin{equation}
\label{eq:geostrophyBoussinesq}
- \khat \times f \surf{\bu_g} = \rBr \surf{\nabla_z p} = \rBr \nabla \pn + g \rBr \rn \nabla \zn = \nabla \Ps.
\end{equation}
Now, the task of a \gsf is to bring $\rn$ inside the gradient with $\zn$. It is easier to work with $\rho$ rather than $\sv$. 
Second, the \isd (still used in the vertical momentum equation, perhaps reduced to hydrostatic balance) is calculated from a Boussinesq equation of state $\eosB$ that uses depth $z$ rather than the \emph{in-situ} pressure \citep{young2010}. Specifically,
\begin{equation}
\rho = \eosB(S, \theta, z) = \frac{1}{\eossv(S, \theta, -g \rB z)}.
 \end{equation}
Thus the neutral surface relation \eqref{eq:def_ntp_svp} becomes
\begin{equation}
\label{eq:def_ntp_rz}
 \nabla \rn = \rzn \nabla \zn.
\end{equation} 
where $\rho_z = \di_z B(S,\theta,z)$.
In the seawater Boussinesq approximation, the multivalued function of pressure, $\svfn$, gets replaced by a multivalued function of depth, $\rfn$. The essential ideas, though, are unchanged. See the appendices for Boussinesq formulas and their numerical discretisation. 

\section{Numeric comparison of geostrophic velocity errors}
\label{sec:results}

The exact \gsf for neutral surfaces, $\Psn$, is inexact on approximately neutral surfaces.  Also, $\Pstb$ differs slightly from $\Psn$ because of the requirement to be well-defined, and because the form \eqref{eq:svfn_form} of the empirically fit functions is limiting.  Nonetheless, $\Pstb$ may estimate the geostrophic velocity more accurately than other \gsfs. The closer to neutral the surface, the more likely $\Pstb$ is to outperform other \gsfs. Numerical tests are required to assess the value of $\Pstb$, as well as $\Psom$. 

\subsection{Setup of numerical tests}
\label{sec:setup}
Three types of approximately neutral surfaces are computed:
\begin{enumerate}[(I)]
\item $\sigma_1$- or $\sigma_2$-surfaces, \ie isosurfaces of potential density \citep{wust1935}; 
\item $\tau'$-surfaces, \ie modified topobaric surfaces \citepalias[][and this manuscript]{stanley2019topobaric}; and
\item $\omega$-surfaces \citep{klocker.mcdougall.ea2009}.
\end{enumerate}
Two of each of these surfaces is computed, intersecting ($\xref$, $\yref$) at
\begin{enumerate}[{(i)}]
\item  $\SI{\zrefone}{m}$, and
\item  $\SI{\zreftwo}{m}$.
\end{enumerate}
These are the depths that the $\omega$-surface heaved to from its initial depth at ($\xref$, $\yref$) of $\SI{-1000}{m}$ and $\SI{-2000}{m}$. 
All six surfaces are masked to exclude where any of them rise into the mixed layer\footnote{
The mixed layer depth is taken as the depth, found by linear interpolation, at which potential density referenced to $\SI{100}{dbar}$ equals that at $\SI{-15}{m}$ (the second shallowest grid cell) plus $\SI{0.03}{kg.m^{-3}}$ \citep{dong.sprintall.ea2008}. 
}, where non-neutral dynamics take over. 
The modified topobaric surfaces differ slightly from regular topobaric surfaces; the area-weighted root-mean-square of the fictitious diapycnal diffusivity (\citealt{mcdougall.jackett2005assessment}; \citealt{klocker.mcdougall.ea2009}; \citetalias{stanley2019topobaric}) over the regular topobaric surfaces are
 (i) $\SI{1.16e-5}{m^2.s^{-1}}$ and (ii) $\SI{7.34e-7}{m^2.s^{-1}}$, 
 whereas for modified topobaric surfaces these numbers are
(i) $\SI{1.15e-5}{m^2.s^{-1}}$ and (ii) $\SI{8.69e-7}{m^2.s^{-1}}$.

On each of these six surfaces, the full velocity $\bu$ is vertically interpolated onto each surface:
\begin{enumerate}[(a)]
\item $\surf{\bu}$. 
\end{enumerate}
Then, the geostrophic velocity is estimated from five \gsfs (all modified for the Boussinesq ocean), as well as from the $z$-level gradient of the pressure:
\begin{enumerate}[(a)]
\setcounter{enumi}{1} 
\item $\buzh = f^{-1} \khat \times \nabla \Pszh$, using \eqref{eq:ZH};
\item $\bucu = f^{-1} \khat \times \nabla \Pscu$, using \eqref{eq:CU}; 
\item $\bumk = f^{-1} \khat \times \nabla \Psmk$, using \eqref{eq:MK};
\item $\buom = f^{-1} \khat \times \nabla \Psom$, using \eqref{eq:orthobaric_montgomery};
\item $\butb = f^{-1} \khat \times \nabla \Pstb$, using \eqref{eq:geostrfexact2} while satisfying \eqref{eq:island};
\item $\buzs = f^{-1} \khat \times \rBr \surf{\nabla_z p}$, from  \eqref{eq:geostrophyBoussinesq}.
\end{enumerate}

For each velocity estimate $\bu^{(i)}$ given by (a)--(g), the error 
\begin{equation}
\bm{\epsilon} = \epsilon^{(x)} \ihat + \epsilon^{(y)} \jhat = \bu^{(i)} - \butruth
\end{equation}
is calculated, where, following \eqref{eq:geostrophyBoussinesq}, 
\begin{equation}
\label{eq:ug_insurface}
\butruth = \frac{1}{f} \, \khat \times \left( \frac{1}{\rB} \nabla \pn + g \frac{\rn}{\rB} \nabla \zn \right)
\end{equation}
is taken as the ``true'' geostrophic velocity. 
Because the transformation from \eqref{eq:defstrf} to \eqref{eq:defstrf2a}, \ie transforming $\nabla_z p$ to $\nabla_r p$, is the first step for all \gsfs, $\butruth$ is a more useful ``truth'' than 
$\buzs$. 
But $\buzs$ also has a fair claim to be the true geostrophic velocity, so the difference between $\buzs$ and $\butruth$ represents the precision with which we may know the actual geostrophic velocity.
The actual geostrophic velocity is ambiguous below this precision, and there is little point reducing $\bm{\epsilon}$ below that for (g).

Two metrics of $\bm{\epsilon}$ will be calculated: the area-weighted $\ell_1$ and $\ell_2$ norms, respectively given by
\begin{equation}
\label{eq:l1}
\norm{\epsilon}_1 = \frac{\sum_i A_i |\epsilon_i|}{\sum_i A_i}
\end{equation}
and
\begin{equation}
\label{eq:l2}
\norm{\epsilon}_2 = \sqrt{\frac{\sum_i A_i \epsilon_i^2}{\sum_i A_i}},
\end{equation}
where $\epsilon = [ \epsilon^{(x)}, \epsilon^{(y)}]$ is a 1D array concatenating the zonal and meridional errors within a given mask, and $A_i$ is the area of the grid cell at the location of $\epsilon_i$. 
The mask, which is common for all (a)--(g) on a given surface, excludes regions not connected to the main ocean, and excludes 1\deg on either side of the equator where geostrophy is invalid (also recall the mixed layer has already been excluded).
These metrics serve for comparing \gsfs in Section~\ref{sec:results_metrics}, but first Section~\ref{sec:results_maps} maps just the zonal geostrophic velocity errors, $\epsilon^{(x)}$. (The meridional errors are qualitatively similar.) The metrics $\norm{\epsilon^{(x)}}_1$ and $\norm{\epsilon^{(x)}}_2$ are defined as in \eqref{eq:l1} and \eqref{eq:l2}, but taking $\epsilon = [\epsilon^{(x)}]$ only. 

\subsection{Mapping geostrophic velocity errors}
\label{sec:results_maps}

\begin{figure*}[!t]
  \centering
  \includegraphics[width=1\textwidth]
{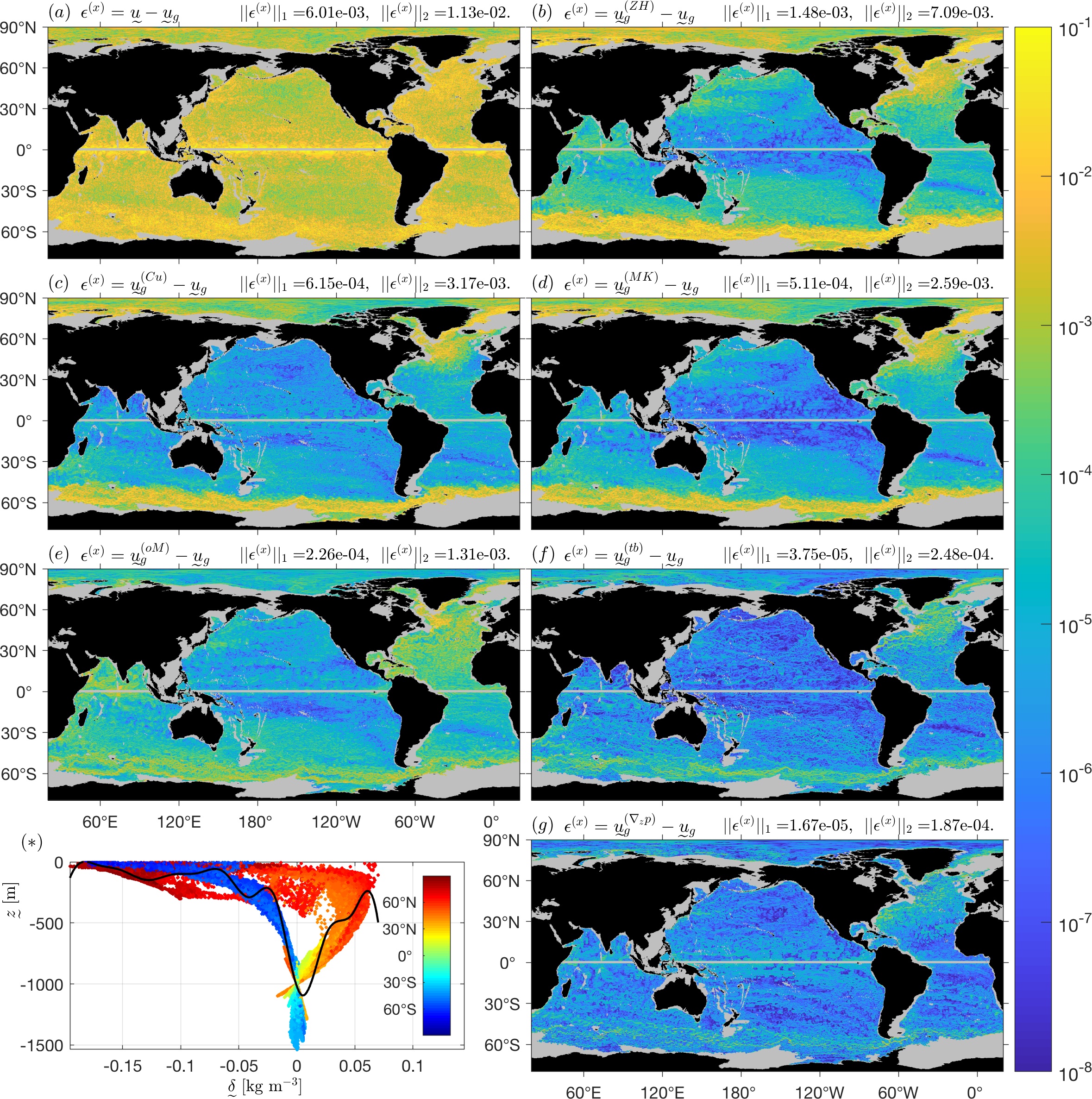}
  \caption{
    Maps of $|\epsilon^{(x)}|$ [$\si{m.s^{-1}}$] the absolute difference between the ``true'' zonal geostrophic velocity $\utruth$ and the zonal velocity 
    from the full velocity (a),
    from five \gsfs (b)--(f),  and
    from $z$-level pressure gradients (g),
     on the $\sigma_1$-surface intersecting ($\xref$, $\yref$, $\SI{\zrefone}{m}$).
  The area-weighted $\ell_1$ and $\ell_2$ norms of $\epsilon^{(x)}$ are listed above each panel, taken inside a mask that excludes (grey regions) the mixed layer, regions that are disconnected from the main ocean, and 1\deg on either side of the equator.
Finally, $\zn$ \vs $\dn$ on this surface and within the same mask are plotted with colour indicating latitude $(*)$, together with the single-valued function used to evaluate $\Psom$ (black curve). 
  }
\label{fig:maps_sigma_1000}
\end{figure*}

\begin{figure*}[!bt]
  \centering
  \includegraphics[width=1\textwidth]
{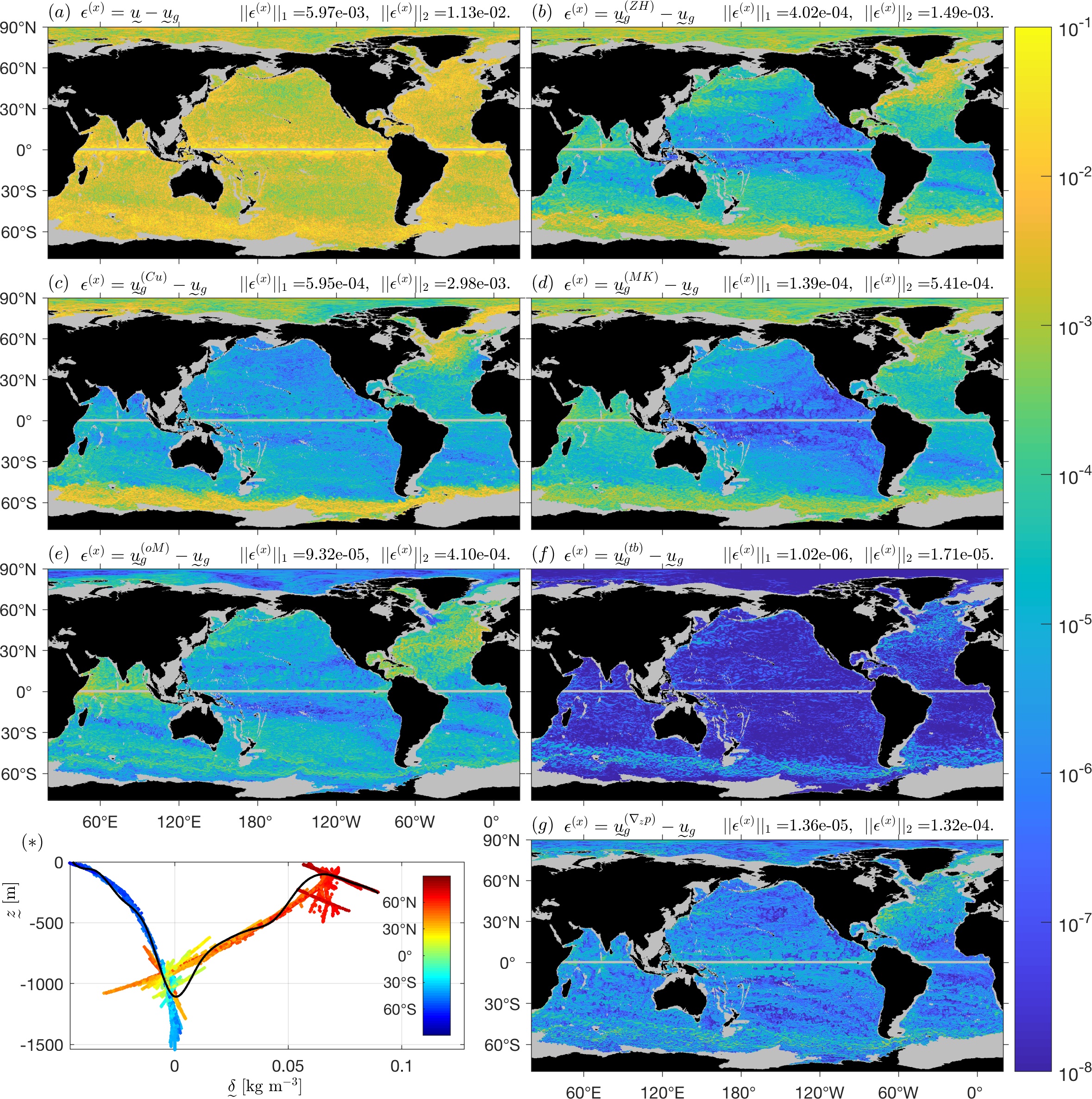}
  \caption{As in Fig.~\ref{fig:maps_sigma_1000} but on the $\tau'$-surface intersecting ($\xref$, $\yref$, $\SI{\zrefone}{m}$).
    }
\label{fig:maps_tau_1000}
\end{figure*}

\begin{figure*}[!t]
  \centering
  \includegraphics[width=1\textwidth]
{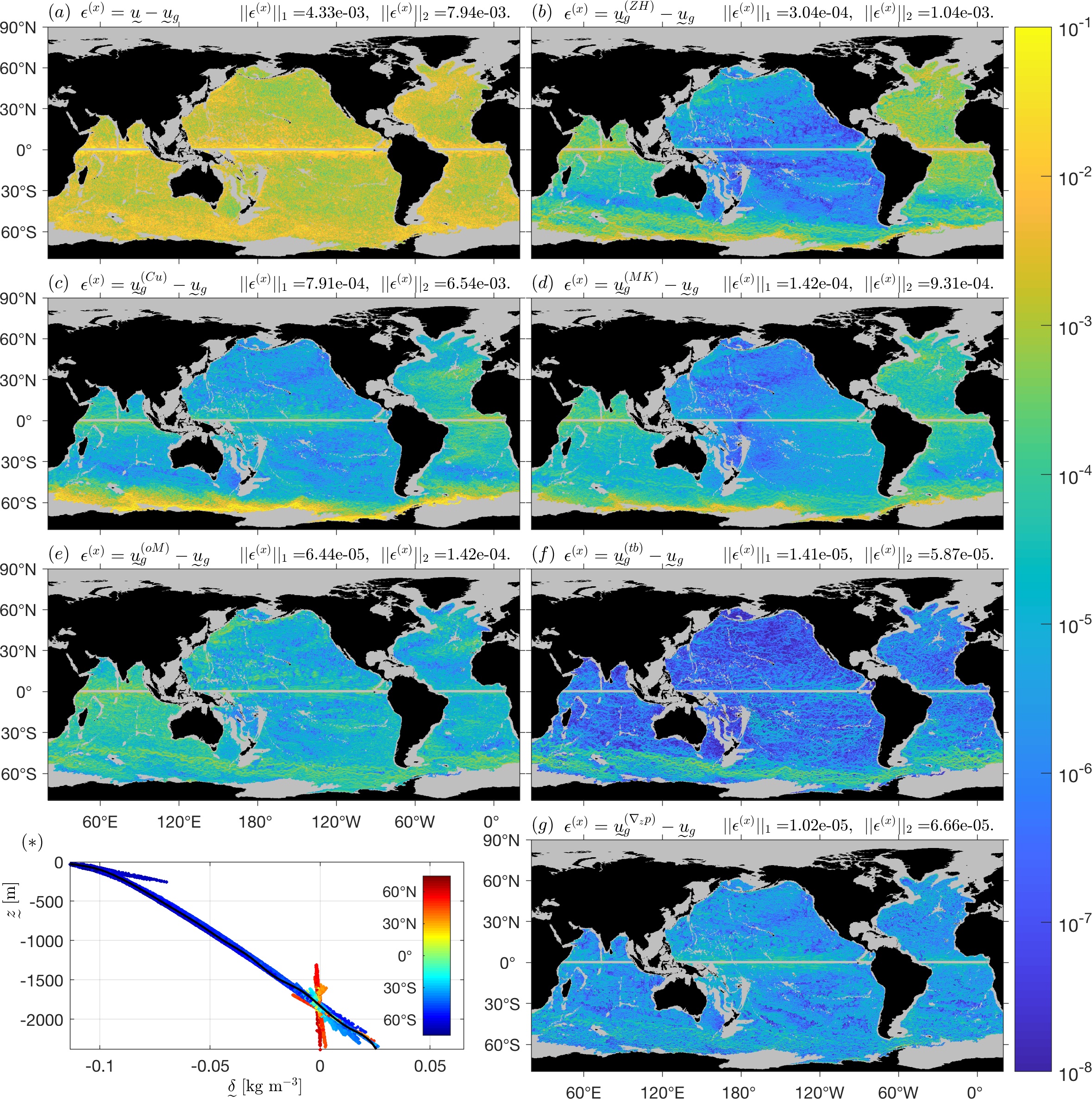}
  \caption{
  As in Fig.~\ref{fig:maps_sigma_1000} but on the $\omega$-surface intersecting ($\xref$, $\yref$, $\SI{\zreftwo}{m}$).  
  }
\label{fig:maps_omega_2000}
\end{figure*}

Figures~\ref{fig:maps_sigma_1000}--\ref{fig:maps_omega_2000} map $\epsilon^{(x)}$ on different surfaces. 
First consider Fig.~\ref{fig:maps_sigma_1000}(b)--(f), on the $\sigma_1$-surface.
$\Pszh$ and $\Psmk$ both perform well in the Indo-Pacific and the subtropical Atlantic, estimating $\butruth$ with accuracy generally below $\mathcal{O}(\SI{1}{mm.s^{-1}})$. 
Here, $\sn$, $\tn$, and $\zn$ are near their reference values, so errors are small --- recall \eqref{eq:ZH}. 
In the North Atlantic/Arctic and Southern Ocean, $\Pszh$ and (to a slightly lesser extent) $\Psmk$ perform poorly, estimating $\butruth$ with accuracy of $\mathcal{O}(\SI{10}{mm.s^{-1}})$, because $\dn$ varies considerably in these regions as the $\sigma_1$-surface shoals up to about $\SI{-100}{m}$ and to the sea surface, respectively. 
For $\Pscu$, errors are small again in the Indo-Pacific and subtropical Atlantic where the surface remains within about $\SI{50}{m}$ of its reference depth of $\SI{\zrefone}{m}$, but errors are large again in the North Atlantic/Arctic and Southern Ocean, both because of large gradients of $\sn$ and $\tn$ here and because the surface is far from the reference depth --- recall \eqref{eq:CU_grad}.
For $\Psom$, the reference depth is a function of $\delta$,
but the $\sigma_1$-surface is far from neutral, so this functional relationship is not very tight, evident by the significant scatter between $\zn$ and $\dn$ in Fig.~\ref{fig:maps_sigma_1000}$(*)$. 
Again, $\Psom$ exhibits its largest errors in the North Atlantic  and Southern Ocean, and to some degree in the Indian ocean, though it performs well in the Arctic; overall $\Psom$ estimates $\butruth$ more accurately than do $\Pszh$, $\Pscu$, or $\Psmk$. 
Being geographically dependent, $\Pstb$ keeps errors low, globally, estimating $\butruth$ with accuracy generally below $\mathcal{O}(\SI{0.1}{mm.s^{-1}})$. Its highest errors are also found in the North Atlantic/Arctic, and Southern Ocean, where the $\sigma_1$-surface is furthest from neutral. 

What level of geostrophic velocity error should be deemed acceptable? The ageostrophic zonal speed (Fig.~\ref{fig:maps_sigma_1000}a) is typically in the range \SIrange{1}{100}{mm.s^{-1}}; estimates of the geostrophic velocity should be at least as accurate as this. Most \gsfs tested here pass this test in most places, but $\Pszh$, $\Pscu$, and $\Psmk$ fail this test in the North Atlantic/Arctic and Southern Ocean. 
On the other hand, it is unreasonable to wish the geostrophic velocity error be less than the difference between calculating the geostrophic velocity on $z$-levels or in the $\sigma_1$-surface (Fig.~\ref{fig:maps_sigma_1000}h). 
Only $\Pstb$ pushes close to this threshold of precision essentially everywhere.

Now, consider Fig.~\ref{fig:maps_tau_1000}, on the upper $\tau'$-surface. 
The errors in estimating the geostrophic velocity from all \gsfs (b)--(f) are smaller on this $\tau'$-surface than on the $\sigma_1$-surface,
because the $\tau'$-surface is closer to neutral. 
This reduces errors for $\Psom$ and $\Pstb$ because their empirically fit functions become more accurate: notice the reduced scatter of $\zn$ \vs $\dn$ in Fig.~\ref{fig:maps_tau_1000}$(*)$. In fact, $\tau'$-surfaces make this relationship exact: there is actually zero scatter between $\zn$ \vs $\dn$, and the apparent scatter in Fig.~\ref{fig:maps_tau_1000}$(*)$ is actually an inability to distinguish between thousands of regions (arcs of the Reeb graph) at the scale shown. 
As for $\Pszh$ and $\Psmk$, the better neutrality of the $\tau'$-surface tends to reduce $|\nabla \dn|$. To see this, expand $\nabla \dn = \surf{\sv_S} \nabla \sn + \surf{\sv_\theta} \nabla \tn + \big( \svpn - \di_p \eossv(S_\rvm, \theta_\rvm, \pn) \big) \nabla \pn$ (reverting to the non-Boussinesq form for familiarity), then note the first two terms are close to zero for nearly neutral surfaces, by \eqref{eq:def_ntp_st}. 

Finally, consider Fig.~\ref{fig:maps_omega_2000}, on the lower $\omega$-surface. 
Again, $\Pszh$, $\Pscu$, and $\Psmk$ perform well in the Pacific where the surface properties remain near the reference properties.
In the subtropical Atlantic though, $\zn$ is about $\SI{250}{m}$ shallower than in the Pacific, and this depth difference increases errors for $\Pscu$ over most of the subtropical Atlantic. Also, this surface samples North Atlantic Deep Water, so $\sn$ and $\tn$, and hence $\dn$, in the Atlantic differ substantially from their Pacific reference values, which causes considerable errors for $\Pszh$ across the entire Atlantic; $\Psmk$ is similarly affected but to a lesser degree.
Once again, $\Pszh$, $\Pscu$, and $\Psmk$ exhibit large errors in the Southern Ocean, particularly near the outcrop where the surface is farthest from its reference depth, and the surface properties furthest from their reference values. 
As $\omega$-surfaces are extremely neutral, $\zn$ is tightly related to $\dn$, so $\Psom$ keeps its reference depth close to $\zn$, and $\Psom$ performs well, globally. 
Again, $\Pstb$ produces the best estimate of $\butruth$, keeping errors low, globally.

For most \gsfs, $\bm{\epsilon}$ is high where the geostrophic velocity itself is high, particularly in the Southern Ocean and North Atlantic, and to a lesser degree in the Kuroshio current and near the equator. 
We could instead study the relative error (divide $\bm{\epsilon}$ by $|\butruth|$), but this creates the opposite problem: errors tend to be high where $\butruth$ is near zero. Nonetheless, the results are qualitatively the same for relative and absolute errors (not shown), so we proceed with the latter.

\subsection{Quantitative error measurements}
\label{sec:results_metrics}

\begin{figure*}[!tb]
  \centering
  \includegraphics[width=1\textwidth]
{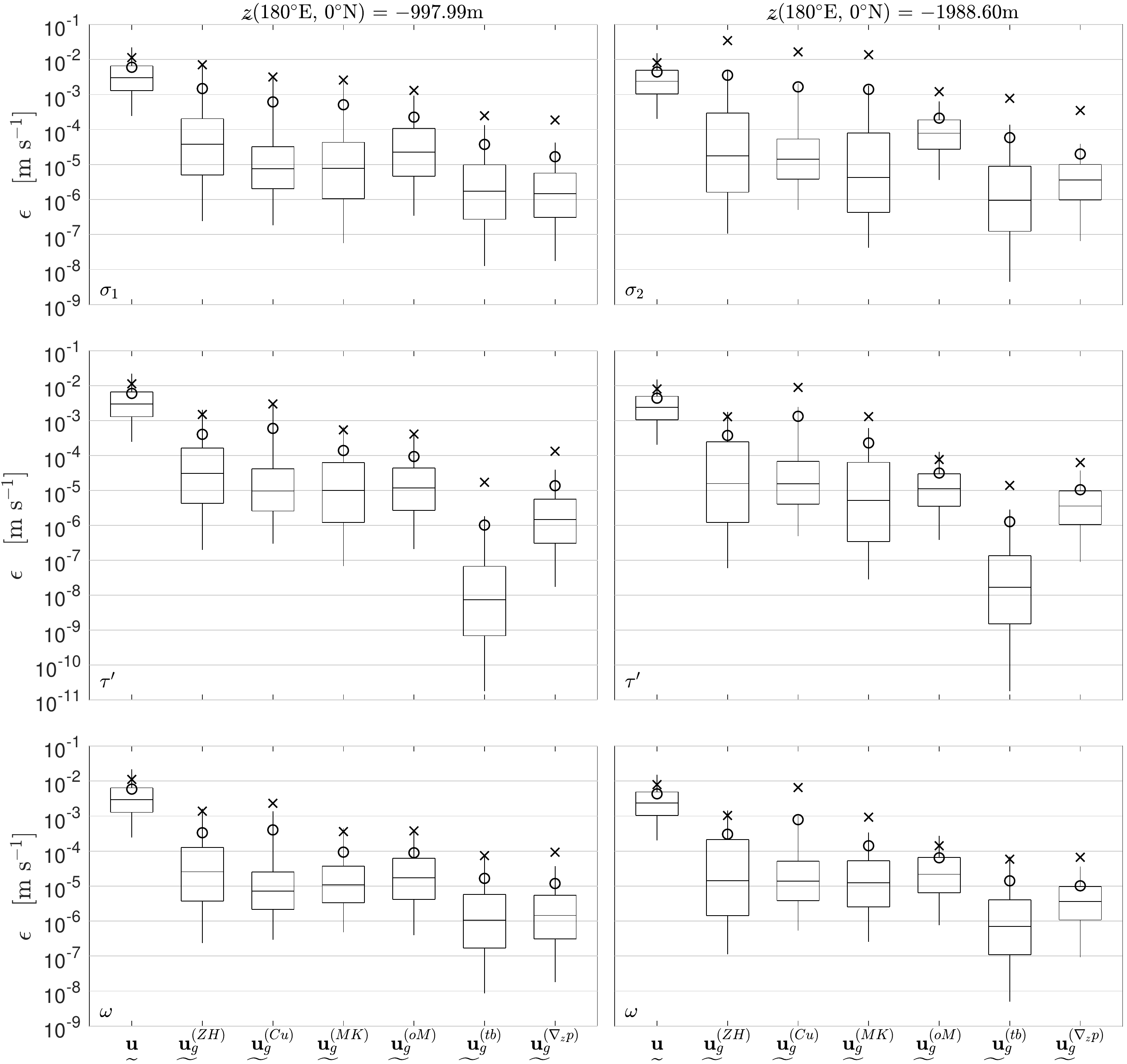}
  \caption{
Norms and boxplots of $\epsilon$.
The six panels show different surfaces, all of which ignore the mixed layer and 1\deg on either side of the equator: 
potential density surfaces (top), 
$\tau'$-surfaces (middle), and
$\omega$-surfaces (bottom)
that intersect 
($\xref$, $\yref$) at 
$\SI{\zrefone}{m}$ (left) or 
$\SI{\zreftwo}{m}$ (right). 
Each panel provides the .05, .25, .5, .75, and .95 area-weighted quantiles of $|\epsilon|$ (boxplots), $\norm{\epsilon}_1$ (circles), and $\norm{\epsilon}_2$ (crosses), 
where $\epsilon$ collects the zonal and meridional components of the error in estimating the geostrophic velocity 
by the full velocity (leftmost),
by $z$-level pressure gradients (rightmost), and
by \gsfs (middle five).
  }
\label{fig:norms_boxplots}
\end{figure*}

We now turn to more quantitative comparisons of the geostrophic velocity errors. 
Figure~\ref{fig:norms_boxplots} shows $\norm{\epsilon}_1$ and $\norm{\epsilon}_2$, as well as boxplots providing the area-weighted 0.05, 0.25, 0.5, 0.75, and 0.95 quantiles of $\epsilon$  for the seven velocity estimates on the three surface classes at both depths.  For numeric clarity, $\norm{\epsilon}_1$ and $\norm{\epsilon}_2$ are also listed in Table~\ref{tab:norms}.  The following discussion focuses mostly on $\norm{\epsilon}_2$.

For $\Pszh$, $\norm{\epsilon}_2$ is about \SIrange{1}{1.5}{mm.s^{-1}} on the $\tau'$- and $\omega$-surfaces, but about \SIrange{7}{35}{mm.s^{-1}} on the $\sigma_1$- and $\sigma_2$-surfaces. 
These errors are reduced for $\Psmk$ by a factor of about 2.5 on the $\sigma_1$- and $\sigma_2$-surfaces, whereas on the $\tau'$- and $\omega$-surfaces this factor is about \numrange{3}{4} on the upper surfaces and about 1 (no reduction) on the lower surfaces. 
For the orthobaric Montgomery potential $\Psom$ 
this factor is roughly \numrange{3.5}{5.5} on the upper surfaces, and \numrange{7}{29} on the lower surfaces. 
For the topobaric geostrophic streamfunction, $\Pstb$, this reduction factor is in the range \numrange{17}{45} on the non-$\tau'$-surfaces (it is hardly fair to compare against $\Pstb$ on the $\tau'$-surface where it is exact).
Indeed, $\Pstb$ improves upon previous results across all measures shown in Fig.~\ref{fig:norms_boxplots}.

The worst performance by $\Pstb$ is on the $\sigma_1$- and $\sigma_2$-surfaces, which are furthest from neutral. On $\omega$-surfaces, $\Pstb$ estimates $\butruth$ about as well as does $\buzs$. 
On the $\tau'$-surfaces, $\butb$ exhibits $\norm{\epsilon}_1$ and $\norm{\epsilon}_2$ about an order of magnitude smaller than those for $\buzs$. In fact, $\Pstb$ is exact on $\tau'$-surfaces, and $\butb - \butruth$ reveals errors not in $\butb$ but in $\butruth$. 
Specifically, $\butruth$ involves a finite difference of $\zn$ multiplied by a simple average of $\rn$, whereas $\butb$ involves a finite difference of two integrals of $\rfn$ with respect to depth. These would be equivalent if $\rfn$ were an affine linear function of $\zn$, but this is false for two reasons.
First, $\rfn$ is multivalued, and this matters when the finite difference selects points in different domains of the branches of $\rfn$.
Second, each branch of $\rfn$ is, by our choice \eqref{eq:svfn_form}, a quadratic function of depth (from integrating an empirically fit affine linear function) plus a rational function of depth (from the equation of state). 

Table~\ref{tab:area_gt_ageo} lists the percentage of surface area within which the estimated geostrophic velocity exceeds (in magnitude) the ageostrophic velocity. For $\Pszh$, this area occupies almost 7\% of the $\sigma_1$-surface, while $\Pscu$ and $\Psmk$ reduce this area to about 3\%, and $\Psom$ reduces it to about 2\%. For $\Pstb$, this area is a mere 0.3\%. Results are similar on the other five surfaces. 
It is therefore important to use an accurate \gsf, particularly when working on global surfaces, lest the ageostrophic velocity be swamped by errors in estimating the geostrophic velocity. 

From a numerical perspective, the trouble caused by the ill-defined nature of the exact \gsf for neutral surfaces tends to be rather small.
For example, on the $\omega$-surface intersecting ($\xref$, $\yref$, $\SI{\zreftwo}{m}$), two topobaric \gsfs are estimated using the methods of Section~\ref{sec:topobaric_gsf}b but excluding the cycle constraints \eqref{eq:island}. They differ only in that they integrate $\svfn_a$ differently on each arc $a$ that defines a cycle in the cycle basis: one integrates up from $p_{l_a}$ and the other down from $p_{h_a}$. 
The geostrophic velocity estimated by these two \gsfs differs at 41,140 grid points, out of a total of 868,954 grid points on this surface (4.7\%). The (unweighted) $\ell_2$ norm of the velocity difference over the grid points where they differ at all is a mere $\SI{3.6e-4}{m.s^{-1}}$.

Finally, the orthobaric Montgomery potential is indeed more accurate than the orthobaric \gsf. 
On the $\omega$-surface through ($\xref$, $\yref$, $\SI{\zreftwo}{m}$), $\norm{\epsilon}_2 = \SI{1.4e-4}{m.s^{-1}}$ for $\Psom$, whereas for the orthobaric \gsf (fitting $\dn$ to $\pn$ as a cubic spline with 12 pieces), $\norm{\epsilon}_2 = \SI{2.7e-4}{m.s^{-1}}$. 
The utility of the orthobaric Montgomery potential becomes more apparent for surfaces that include both the Southern Ocean and the Arctic: on the $\omega$-surface through ($\xref$, $\yref$, $\SI{\zrefone}{m}$), $\norm{\epsilon}_2 = \SI{3.8e-4}{m.s^{-1}}$ for $\Psom$, whereas for the orthobaric \gsf, $\norm{\epsilon}_2 = \SI{9.2e-4}{m.s^{-1}}$. 

\begin{table}[!t]
\tabcolsep=0.1cm
\begin{center}
\begin{tabular}{|l|ccc|ccc|}
\hline
 (a) & \multicolumn{3}{c|}{$\zn(\bm{x}_\rv) = \SI{-997.99}{m}$} & \multicolumn{3}{c|}{$\zn(\bm{x}_\rv) = \SI{-1988.60}{m}$} \\ 
$||\mathbf{\epsilon}||_1$& $\sigma_{1}$ & $\tau'$ & $\omega$ & $\sigma_{2}$ & $\tau'$ & $\omega$ \\[5pt]
\hline
& & & & & & \\[-7pt]
$\utilde{\mathbf{u}}$ & 6.0e-03 & 6.0e-03 & 5.9e-03 & 4.4e-03 & 4.4e-03 & 4.3e-03 \\[5pt]
$\utilde{\mathbf{u}_g}^{(ZH)}$ & 1.5e-03 & 4.0e-04 & 3.3e-04 & 3.6e-03 & 3.7e-04 & 3.0e-04 \\[5pt]
$\utilde{\mathbf{u}_g}^{(Cu)}$ & 6.2e-04 & 6.0e-04 & 4.0e-04 & 1.7e-03 & 1.3e-03 & 7.9e-04 \\[5pt]
$\utilde{\mathbf{u}_g}^{(MK)}$ & 5.1e-04 & 1.4e-04 & 9.3e-05 & 1.4e-03 & 2.3e-04 & 1.4e-04 \\[5pt]
$\utilde{\mathbf{u}_g}^{(oM)}$ & 2.3e-04 & 9.3e-05 & 8.9e-05 & 2.1e-04 & 3.1e-05 & 6.4e-05 \\[5pt]
$\utilde{\mathbf{u}_g}^{(tb)}$ & 3.8e-05 & 1.0e-06 & 1.7e-05 & 5.9e-05 & 1.3e-06 & 1.4e-05 \\[5pt]
$\utilde{\mathbf{u}_g}^{(\nabla_z p)}$ & 1.7e-05 & 1.4e-05 & 1.2e-05 & 2.0e-05 & 1.0e-05 & 1.0e-05 \\[5pt]
\hline
\end{tabular}
\end{center}
\begin{center}
\begin{tabular}{|l|ccc|ccc|}
\hline
 (b) & \multicolumn{3}{c|}{$\zn(\bm{x}_\rv) = \SI{-997.99}{m}$} & \multicolumn{3}{c|}{$\zn(\bm{x}_\rv) = \SI{-1988.60}{m}$} \\ 
$||\mathbf{\epsilon}||_2$& $\sigma_{1}$ & $\tau'$ & $\omega$ & $\sigma_{2}$ & $\tau'$ & $\omega$ \\[5pt]
\hline
& & & & & & \\[-7pt]
$\utilde{\mathbf{u}}$ & 1.1e-02 & 1.1e-02 & 1.1e-02 & 8.1e-03 & 8.0e-03 & 7.9e-03 \\[5pt]
$\utilde{\mathbf{u}_g}^{(ZH)}$ & 7.1e-03 & 1.5e-03 & 1.4e-03 & 3.5e-02 & 1.3e-03 & 1.0e-03 \\[5pt]
$\utilde{\mathbf{u}_g}^{(Cu)}$ & 3.2e-03 & 3.0e-03 & 2.3e-03 & 1.7e-02 & 8.9e-03 & 6.5e-03 \\[5pt]
$\utilde{\mathbf{u}_g}^{(MK)}$ & 2.6e-03 & 5.4e-04 & 3.6e-04 & 1.4e-02 & 1.3e-03 & 9.3e-04 \\[5pt]
$\utilde{\mathbf{u}_g}^{(oM)}$ & 1.3e-03 & 4.1e-04 & 3.8e-04 & 1.2e-03 & 7.6e-05 & 1.4e-04 \\[5pt]
$\utilde{\mathbf{u}_g}^{(tb)}$ & 2.5e-04 & 1.7e-05 & 7.4e-05 & 7.8e-04 & 1.4e-05 & 5.9e-05 \\[5pt]
$\utilde{\mathbf{u}_g}^{(\nabla_z p)}$ & 1.9e-04 & 1.3e-04 & 9.3e-05 & 3.5e-04 & 6.3e-05 & 6.7e-05 \\[5pt]
\hline
\end{tabular}
\end{center}
\caption{
The $\ell_1$ (a) and $\ell_2$ (b) norms of the error $\epsilon$ [$\si{m.s^{-1}}$] in estimating the ``true'' geostrophic velocity, for various velocity estimates (rows) and surfaces (columns). In the top row, the velocity estimate is the full velocity, so $\epsilon$ is the ageostrophic velocity. 
}
\label{tab:norms}
\end{table}

\begin{table}[!t]
\tabcolsep=0.28cm
\begin{center}
\begin{tabular}{|l|ccc|ccc|}
\hline
& \multicolumn{3}{c|}{$\zn(\bm{x}_\rv) = \SI{-997.99}{m}$} & \multicolumn{3}{c|}{$\zn(\bm{x}_\rv) = \SI{-1988.60}{m}$} \\ 
& $\sigma_{1}$ & $\tau'$ & $\omega$ & $\sigma_{2}$ & $\tau'$ & $\omega$ \\[5pt]
\hline
& & & & & & \\[-7pt]
$\utilde{\mathbf{u}_g}^{(ZH)}$ & 6.90 & 3.30 & 2.50 & 7.91 & 5.05 & 4.25 \\[5pt]
$\utilde{\mathbf{u}_g}^{(Cu)}$ & 3.41 & 3.39 & 2.04 & 4.31 & 4.84 & 3.21 \\[5pt]
$\utilde{\mathbf{u}_g}^{(MK)}$ & 3.07 & 1.32 & 0.71 & 4.24 & 2.34 & 1.40 \\[5pt]
$\utilde{\mathbf{u}_g}^{(oM)}$ & 2.00 & 1.08 & 1.09 & 3.11 & 0.48 & 1.02 \\[5pt]
$\utilde{\mathbf{u}_g}^{(tb)}$ & 0.28 & 0.00 & 0.14 & 0.47 & 0.01 & 0.15 \\[5pt]
$\utilde{\mathbf{u}_g}^{(\nabla_z p)}$ & 0.12 & 0.11 & 0.10 & 0.19 & 0.15 & 0.15 \\[5pt]
\hline
\end{tabular}
\end{center}
\caption{The percentage of the surface area, on different surfaces (columns), for which the error in estimating the ``true'' geostrophic velocity by methods (b)--(g) (rows) exceeds the ageostrophic velocity, in magnitude.}
\label{tab:area_gt_ageo}
\end{table}

\section{Summary}
\label{sec:conclusions}

The exact \gsf on a neutral surface, whose existence has long been known \citep{mcdougall1989}, has been derived. It is defined using path integrals, along neutral trajectories, of the specific volume as a function of pressure.  On a (hypothetical) well-defined neutral surface, the specific volume is a multivalued function of the pressure, and its geographic structure is described by the \citet{reeb1946} graph of the pressure on the surface \citepalias{stanley2019topobaric}. The path integrals defining the \gsf are equivalently described by a sum of simple integrals determined by a walk through the Reeb graph. 
Islands, and other holes in the neutral surface, can create cycles in the Reeb graph that may be walked in either direction. That is, they cause the path integral to be path-dependent, and the exact \gsf for a neutral surface is actually a multivalued function of geographic position.
When mapping the \gsf, its multivalued nature appears as discontinuities emanating from islands and other such holes. 
Though the gradient of the \gsf, and hence the geostrophic velocity, is unaffected by this multivalued nature, it does hamper the use of streamlines for flow visualization or other analyses.

The problem of these discontinuities is overcome by the topobaric \gsf, which everywhere approximates the exact \gsf for neutral surfaces, but is well-defined. Moreover, the topobaric \gsf can be used on any approximately neutral surface. On a general approximately neutral surface, the specific volume is empirically fit as a multivalued function of pressure, subject to a set of constraints determined by cycles in the Reeb graph, that ensure the topobaric \gsf is well-defined. Numerical tests reveal that the topobaric \gsf estimates the geostrophic velocity on potential density surfaces and $\omega$-surfaces far more accurately than any other known \gsf.  On topobaric surfaces, the specific volume is given \emph{a priori} as a multivalued function of pressure.  Numerical tests confirm that the topobaric \gsf is exact on modified topobaric surfaces. 

The \citet{montgomery1937} potential has also been generalized, furthering the progress made by \citet{zhang.hogg1992}. This leads to an alternative formulation of the exact \gsf for neutral surfaces, which is amenable to an approximation wherein the pressure on a surface is empirically fit as a single-valued function of the specific volume anomaly on the surface. 
This approximation, called the orthobaric Montgomery potential, is easy to compute and improves upon all previously known \gsfs for approximately neutral surfaces.

\section*{Acknowledgements}
The author thanks David Marshall for helpful discussions, 
Trevor McDougall for early encouragement to pursue this topic, 
two reviewers for their constructive feedback,
and authors of quality software, including Andreas Klocker ($\omega$-surfaces), David Gleich (GAIMC), Jonas Lundgren (splinefit), and Harish Doraiswamy and Vijay Natarajan (ReCon).
GJS was supported by the Clarendon Scholarship, and the Canadian Alumni Scholarship at Linacre College, University of Oxford.
MATLAB software to compute the topobaric geostrophic streamfunction, the orthobaric Montgomery potential, and other results herein is available from the author's website. 

\appendix
\section{Boussinesq geostrophic streamfunctions}
\label{sec:geostrf_boussinesq}

As discussed in Section~\ref{sec:Boussinesq_models}, the Boussinesq approximation swaps the roles of $p$ and $z$, and of $\sv$ and $\rho$. It is helpful to re-define $\delta$ as the \isd anomaly,
\begin{equation}
\label{eq:deltaBSQ}
\delta = \rho - \rho_\rv
\end{equation}
where $\rho_\rv = \eosB(S_\rv, \theta_\rv, z)$ is the \isd at the local depth but at a reference salinity $S_\rv$ and reference potential temperature $\theta_\rv$. 
The derivations of Sections~\ref{sec:theory}--\ref{sec:orthobaric_montgomery} proceed, in essence, unchanged. The results are as follows.

Hydrostatic balance is integrated to obtain the pressure, 
\begin{equation}
\label{eq:hydrostatic_pressure}
\pn = \slp + g \int^\eta_{\zn} \rho \, \dz.
\end{equation}

The exact \gsf for a neutral surface \eqref{eq:geostrfexact} becomes,
\begin{equation}
\label{eq:geostrfexactBSQ}
\Psn = \frac{1}{\rB} \pn - \frac{g}{\rB} \int_{\zn}^{z_\rv} \rfn (z) \, \d z,
\end{equation}
and the graph integral form \eqref{eq:geostrfexact2} transforms similarly. 

For the Montgomery potential, \eqref{eq:montgomery} becomes 
\begin{align}
\label{eq:montgomery_bsq}
\nabla \Psd &= 
\frac{g}{\rB} \zn \nabla \dn \\
&+ \nabla \left( 
\frac{g}{\rB} \zn \ \dn 
- \frac{g}{\rB} \int_{\zn}^{Z} \rho_\rv \, \d z 
+ \frac{1}{\rB} \pn \right) \nonumber
\end{align}
for some constant depth $Z$.

For the \citet{zhang.hogg1992} \gsf, \eqref{eq:ZH} becomes
\begin{align}
\nabla \Pszh &= 
\frac{g}{\rB} (\zn - z_\rv) \nabla \dn \\
& +\nabla \left( 
  \frac{g}{\rB} (\zn - z_\rv) \dn 
- \frac{g}{\rB} \int_{\zn}^{Z} \rho_\rv \, \d z 
+ \frac{1}{\rB} \pn 
\right), \nonumber
\end{align}
for another constant $z_0$. 

For the orthobaric Montgomery potential, \eqref{eq:orthobaric_montgomery} becomes
\begin{align}
&\nabla \Psom = 
\frac{g}{\rB} \big( \zn - \hat{z}(\dn) \big) \nabla \dn  \\
&+ \nabla \left( 
  \frac{g}{\rB} \zn \ \dn 
- \frac{g}{\rB} \int_\Delta^{\dn} \hat{z}(\delta) \, \d \delta 
- \frac{g}{\rB} \int_{\zn}^{Z} \rho_\rv \, \d z 
+ \frac{1}{\rB} \pn 
\right) \nonumber
\end{align}
where $\hat{z}$ is a function of $\delta$.

For the \citet{cunningham2000} \gsf, \eqref{eq:CU} becomes 
\begin{equation}
\label{eq:CU_BSQ}
\Pscu 
= -\frac{g}{\rB} \int^Z_{\zn} \eosB(\sn,\tn,z') \, \d z' + \frac{1}{\rB} \pn.
\end{equation}

For the \citet{mcdougall.klocker2010} \gsf, the Boussinesq form is most easily found by noting how terms transform in the Montgomery potential: $g \zn$ transforms to $\rBr \pn$, and $\pn (\svn - \sv_\rv)$ to $g \rBr \zn (\rn - \rho_\rv)$. 
One can use these rules to determine how the $T_b / (12 \, \rn)$ term in $\Psmk$ transforms to the Boussinesq form, pivoting on their \eqn{55}. The Boussinesq version of \eqref{eq:MK} is
\begin{align}
\label{eq:MK_BSQ}
\Psmk &=
\frac{1}{2}\frac{g}{\rB} (\zn - z_\rv) \dn 
- \frac{g}{\rB} \int_{\zn}^{Z} \rho_\rv \, \d z 
+ \frac{1}{\rB} \pn \nonumber \\
&- \frac{1}{12} \frac{T_b}{\rn} (g \rB)^2 (\surf{\theta} 
- \theta_\rv) (\zn - z_\rv)^2 ,
\end{align}
where again $T_b / \rn = \SI{2.7e-15}{K^{-1} Pa^{-2} m^2 s^{-2}}$ is treated as constant.

\section{Numerical methods}
\label{sec:numerics}

The numerical methods to calculate the various \gsfs from discretised model data require some care. They depend, first, on how the geostrophic velocity is calculated. 
The model has discrete depth levels $z_k < 0$ for the centre of each tracer cell, where the salinity $S_{i,j,k}$ and potential temperature $\theta_{i,j,k}$, and thus the  \isd $\rho_{i,j,k}$, are known in each water column $(i,j)$.  
For each water column, the model calculates the pressure $p_{i,j,k}$ at each $z_k$ by integrating hydrostatic balance, \eqref{eq:hydrostatic_pressure}, using $\rB$ as the density between $z = 0$ and the free surface, $\rho_{i,j,1}$ as the density between $z = 0$ and $z = z_1$, and otherwise using trapezoidal integration:
\begin{align}
\label{eq:trapzintp}
p_{i,j,k} = \ &p^{(\eta)}_{i,j} + g \, \rB \, \eta_{i,j} - g \, \rho_{i,j,1} \, z_1 \nonumber \\
&+ \sum_{k'=2}^k g \, \frac{\rho_{i,j,k'-1} + \rho_{i,j,k'}}{2} \, (z_{k'-1} - z_{k'}) .
\end{align}
This must be extended to give the pressure $p_{i,j}(z')$ at an arbitrary depth $z'$.
Linearly interpolating the pressure implies, via hydrostatic balance, that the \isd is uniformly $(\rho_{i,j,k} + \rho_{i,j,k+1}) / 2$ between $z_k$ and $z_{k+1}$. Then $\rn$ would discontinuously jump where the surface crosses grid points --- most unconscionable --- and so too would the geostrophic velocity given by \eqref{eq:ug_insurface}. 
Instead, the \isd is continuously extended to $\rho_{i,j}(z')$, then hydrostatic balance gives the pressure as
\begin{equation}
p_{i,j}(z') = p_{i,j,k} + g \, \frac{\rho_{i,j,k} + \rho_{i,j}(z')}{2} \, (z_k - z'),
\end{equation}
where $k$ is such that $z_k \leq z' < z_{k+1}$. This form ensures the pressure is continuous, \ie $p_{i,j}(z_k) = p_{i,j,k}$. 

Now suppose we define $\rho_{i,j}(z')$ by linear interpolation:
\begin{equation}
\label{eq:lin_interp_rho}
\rho_{i,j}(z') = \frac{(z' - z_{k+1}) \, \rho_{i,j,k} + (z_k - z') \, \rho_{i,j,k+1}}{z_k - z_{k+1}}.
\end{equation}
The discretised version of the middle equation in \eqref{eq:geostrophyBoussinesq} --- rather, its meridional component, up to a constant scaling factor, and suppressing zonal grid indices for brevity ---
is
\begin{align}
\label{eq:discretized_grad}
& p_{j} (\bar{z}_{j}) - p_{j-1}(\bar{z}_{j}) \\
&= p_{j}(\zn_{j}) - p_{j-1}(\zn_{j-1})
+ \frac{g}{2} \big(\rn_{j} + \rn_{j-1} \big) \big( \zn_{j} - \zn_{j-1} \big) \nonumber \\
& \quad - \frac{g}{8} \big(\rho_{j,k} - \rho_{j-1,k} + \rho_{j-1,k+1} - \rho_{j,k+1} \big) \frac{\big(\zn_{j} - \zn_{j-1}\big)^2}{z_k - z_{k+1}}, \nonumber
\end{align}
where $\bar{z}_j = (\zn_{j-1} + \zn_{j}) / 2$. In \eqref{eq:discretized_grad},  $\zn_{j-1}$ and $\zn_{j}$ are assumed to lie in the same range, $[z_k, z_{k+1}]$. (If they are not, the error term differs, but no discontinuities appear.)
Dividing \eqref{eq:discretized_grad} by the meridional distance between water columns, 
the first line is the discretised $z$-level pressure gradient, the second line is the discretised in-surface pressure gradient plus a contribution that undoes the sloping gradient, and the third line is a third-order error. Zonal discretisations are analogous. 

Now, neutral errors on a surface are traditionally calculated using $\nabla \sn$ and $\nabla \tn$, which has an equivalent form using $\nabla \rn$ and $\nabla \zn$ only if $\rn = B(\sn, \tn, \zn)$ --- see the discussion around \eqn{28} of \citetalias{stanley2019topobaric}.
For consistency with this, instead of \eqref{eq:lin_interp_rho}, we define $\rho_{i,j}(z') = B(S_{i,j}(z'), \theta_{i,j}(z'), z')$, then define $S_{i,j}(z')$ and $\theta_{i,j}(z')$ by linear interpolation analogous to \eqref{eq:lin_interp_rho}.
This introduces extra errors to \eqref{eq:discretized_grad} due to non-linearity in $B$, but they are considerably smaller than the discretization error, the third line in \eqref{eq:discretized_grad}.

Whereas the integral of \isd is designed to match the model's pressure, the integral of the reference \isd 
 has the sole purpose of satisfying 
\begin{equation}
\nabla \int_{\zn}^Z \rho_\rv \, \d z = -\rn_\rv \nabla \zn .
\end{equation}
To best satisfy this numerically, $\rho_\rv$ is trapezoidally integrated using a mesh with $\SI{0.1}{m}$ spacing, much finer than the model's $z_k$ levels. (Where the reference $S$ and $\theta$ are not constants, as in $\Pscu$, the model's $z_k$ levels are used.) 

The remaining integrals, namely the affine linear part of $\Pstb$, and the $\hat{p}$ part of $\Psom$, are handled analytically. 

The TEOS-10 routines to calculate $\Pszh$ and $\Psmk$ are not used, for they assume a too-recent equation of state and a non-Boussinesq ocean, whereas ECCO2 uses the Boussinesq form of the \citet{jackett.mcdougall1995} equation of state. Instead, $\Pszh$ and $\Psmk$ are calculated from scratch, using the aforementioned vertical interpolation and integration methods.

ECCO2 uses a C-grid, so the zonal velocity $u$ lives on the west face of a tracer cell, whereas the zonal geostrophic velocity lives on the south face of a tracer cell. To evaluate $\surf{u}$ --- the zonal component of (a) in Section~\ref{sec:setup} --- the 3D $u$ field is brought to the south face by the same weighted average that the model uses when calculating the Coriolis acceleration, then linearly interpolated to $\bar{z}_{i,j}$. The meridional velocity is handled analogously. 

\section*{References}


\end{document}